\documentclass[conference]{IEEEtran} 
\IEEEoverridecommandlockouts
\usepackage{cite}
\usepackage{amsmath,amssymb,amsfonts}
\usepackage{algorithmic}
\usepackage{graphicx}
\usepackage{textcomp}
\usepackage[table,xcdraw]{xcolor}
\usepackage{subcaption}
\usepackage{listings}
\usepackage{url}
\def\BibTeX{{\rm B\kern-.05em{\sc i\kern-.025em b}\kern-.08emT\kern-.1667em\lower.7ex\hbox{E}\kern-.125emX}}

\definecolor{dkgreen}{rgb}{0,0.6,0}
\definecolor{gray}{rgb}{0.5,0.5,0.5}
\definecolor{mauve}{rgb}{0.58,0,0.82}

\begin{document}

\title{Teaching Network Traffic Matrices in an \\ Interactive Game Environment 
\thanks{This material is based upon work supported by the Under Secretary of Defense for Research and Engineering under Air Force Contract No. FA8702-15-D-0001. Any opinions, findings, conclusions or recommendations expressed in this material are those of the author(s) and do not necessarily reflect the views of the Under Secretary of Defense for Research and Engineering. Research was also sponsored by the United States Air Force Research Laboratory and the Department of the Air Force Artificial Intelligence Accelerator and was accomplished under Cooperative Agreement Number FA8750-19-2-1000. The views and conclusions contained in this document are those of the authors and should not be interpreted as representing the official policies, either expressed or implied, of the Department of the Air Force or the U.S. Government. The U.S. Government is authorized to reproduce and distribute reprints for Government purposes notwithstanding any copyright notation herein.  Use of this work is controlled by the human-to-human license listed in Exhibit 3 of https://doi.org/10.48550/arXiv.2306.09267}
}

\author{\IEEEauthorblockN{Chasen Milner$^{1,2}$, Hayden Jananthan$^2$, Jeremy Kepner$^2$, Vijay Gadepally$^2$, Michael Jones$^2$, \\  Peter Michaleas$^2$, Ritesh Patel$^2$, Sandeep Pisharody$^2$, Gabriel Wachman$^2$, Alex Pentland$^2$
\\
\IEEEauthorblockA{$^1$USAF,  $^2$MIT
}}}
\maketitle

\begin{abstract}
The Internet has become a critical domain for modern society that requires ongoing efforts for its improvement and protection.  Network traffic matrices are a powerful tool for understanding and analyzing networks and are broadly taught in online graph theory educational resources.  Network traffic matrix concepts are rarely available in online computer network and cybersecurity educational resources.   To fill this gap, an interactive game environment has been developed to teach the foundations of  traffic matrices to the computer networking community.  The game environment provides a convenient, broadly accessible, delivery mechanism that enables making material available rapidly to a wide audience.  The core architecture of the game is a facility to add new network traffic matrix training modules via an easily editable JSON file.  Using this facility an initial set of  modules were rapidly created  covering: basic traffic matrices, traffic patterns, security/defense/deterrence, a notional cyber attack, a distributed denial-of-service (DDoS) attack, and a variety of graph theory concepts.  The game environment enables delivery in a wide range of contexts to enable rapid feedback and improvement.  The game can be used as a core unit as part of a formal course or as a simple interactive introduction in a presentation.
\end{abstract}


\section{Introduction}

As the Internet  becomes more critical  for modern society this requires increasing efforts for its improvement and protection.   Network traffic matrices (adjacency matrices in the graph theory community) are a powerful tool for understanding and analyzing networks (and graphs) \cite{kepner2011graph, kepner2018mathematics, szarnyas2021lagraph, bader2022massive}.  Matrix based approaches  have become even more powerful with the advent of high-performance analytic software standards like the GraphBLAS \cite{mattson2013standards, kepner2016mathematical, buluc2017design, brock2021introduction} available in many programming environments \cite{davis2019algorithm, cailliau2019redisgraph, pelletier2021graphblas, dorre2021graphblas, brock2022graphblas, aadhithya2023informal, roose2023tensql}.   These technologies have enabled the anonymized real-time analysis of terabit scale networks with trillions of events \cite{jones2022graphblas, jones2023deployment, bergeron2023hypersparse, kepner2021spatial}.

Briefly, an adjacency matrix
$$
  {\bf A}(i,j) = v
$$
indicates that a graph has edge starting at vertex $i$ and ending at vertex $j$ with corresponding value $v$.  A network traffic matrix is an adjacency matrix where the vertices are sources and destinations on a computer network and the weight represents the amount of information being sent from source $i$ to destination $j$.  Typically $v$ is measured in terms of the number of bytes or data packets sent.  Formally, $i$ and $j$ are chosen from pre-fixed initial segments of the positive integers.  In computer networks, other labels for $i$ and $j$ are often used (such as strings) which can be handled with the more general associative array abstraction \cite{kepner2018mathematics}.  In this paper, the term \emph{traffic matrix} is used regardless of whether the sources and destinations are denoted with integers or strings.

Adjacency matrix concepts are generally available in online graph theory education resources.  Examination of relevant courses from  some popular online learning platforms indicates that adjacency matrix concepts are introduced in some online courses \cite{meyer2015mathematics, kepner2020mathematics, fattah2020graph, rhodes2023algorithms, gupta2023graph, gripon2023advanced}.
Video resources are the most common way to present the educational material with text being much less common.  The reliance on video is understandable with how visually motivated graph theory is.

In the network protection domain, adjacency matrices or their corresponding network traffic matrices are rarely taught in online cybersecurity classes.    The Cybersecurity Labs and Resource Knowledge-base, or CLARK, is an ever growing online library of cybersecurity education \cite{219738} that contains over 1500 cyber educational resources encompassing over 25 topic categories. Over 150 educational resources are collected under the \emph{Network Security} topic with three having matrix content. One resource presents adjacency matrices in the context of social network analysis \cite{golbeck2020predictive}, a second uses matrix-matrix multiply as example parallel computing application \cite{golbeck2020map}, and a third represents neural networks as matrices for machine learning \cite{khan2020probabilistic}.  At the time of writing, none of the CLARK cybersecurity resources present network traffic matrix concepts as a tool for protecting computer networks.  This may be in part because available educational resources currently hosted on CLARK requires little beyond high school mathematics.  This is to be expected, as the intended audience is cybersecurity experts and not scientists, engineers, or mathematicians.  The mathematical sophistication of adjacency matrices can be a barrier to the network security audience which may have limited exposure to advanced mathematics.  Interactive game environments can be a powerful tool for distilling sophisticated concepts into an easily digestible format.

Serious games are a type of video game that focus on acquiring  knowledge, often used for professional training and education \cite{https://doi.org/10.1002/hbe2.188}.  Research over the past several decades has shown that bringing serious games into the classroom is increasing in popularity and can enhance learning, make lessons more enjoyable for the student, and increase motivation and engagement \cite{EKIN2023104700}. When looking at how students from different backgrounds benefit from game-based learning it has been shown that that cognitive, motivational, and behavioural outcomes of educational games show a moderate effect compared to conventional teaching methods \cite{doi:10.1080/03057267.2022.2057732}. Game-based education is not new to the cybersecurity field with notable card-based games Backdoors \& Breaches \cite{backdoorsandbreaches} and Riskio \cite{HART2020101827} as examples of effective teaching tools.

Our \emph{Traffic Warehouse} video game is a prototype designed to allow a user to be introduced to various network traffic matrix concepts.  The game has a convenient broadly accessible delivery mechanism that enables making material available more rapidly to the intended audience as compared to other modalities such as static text or video.  Specfically, the game is designed to be extensible by non-game developers so that it is possible to rapidly add new traffic matrix concepts.  Using the online game approach  the material can be made available to anyone in almost any context.

The rest of the paper is organized as follows.  First, the extensible learning module JSON (JavaScript Object Notation) file is described as this is core educator interface to \emph{Traffic Warehouse}.  Next, a brief review of how the Godot and MagicaVoxel technologies were selected for this development effort.  Some illustrative coding examples are provided to illustrate some details of the implementation.  Finally, several learning modules are presented with their corresponding in-game screenshots.  

\section{Extensible Learning Module File}

The key design choice of the \emph{Traffic Warehouse} game was to define the learning modules via easily editable JSON \cite{jsonrfc} files that a non-game developer could use to create new learning modules.  Using this approach a student is able to load a set of JSON files that contain different traffic matrices with associated multiple choice questions.

Learning modules consist of a zip file containing multiple JSON files that the user can select and load into the game.  \emph{Traffic Warehouse} will take the zip file and load each of the JSON files contained in it and present them sequentially one at a time.  To create a single matrix lesson there are example files that can be duplicated and modified.  Given the core concept being illustrated is the notion of traffic matrices, a key feature is the size of the matrix.  There are template JSON files for $6{\times}6$ or $10{\times}10$ matrices.  Inside the JSON file the first fields are for the title of the lesson, the author, and the traffic matrix axis labels. There is  a single list of axis labels that will be applied to both the vertical and horizontal axis.  Shorter all caps labels are easier to view in the game.  In this example, the labels correspond to work station (WS), server (SRV), external (EXT), and adversary (ADV).

\lstset{language=Java}
\begin{lstlisting}[basicstyle=\ttfamily\scriptsize]
	"name":"10x10 Template",
	"size":"10x10",
	"author":"Chasen Milner",
	"axis_labels":[
		"WS1","WS2","WS3","SRV1",
		"EXT1","EXT2",
		"ADV1","ADV2","ADV3","ADV4",
		],
\end{lstlisting}

The next---and perhaps most critical field in the JSON file---specifies the network traffic matrix.  With this field the number of packets sent between each source and destination is specified.  While there is no hard limit in code, through testing it has been found that fewer than 15 packets between any source and destination displays well.   The matrix is represented as a list of lists to make it intuitive for an educator to type out exactly what the student will see.

\begin{lstlisting}[basicstyle=\ttfamily\scriptsize]
	"traffic_matrix":[
		[1,0,0,0,0,0,0,0,0,2],
		[0,1,0,0,0,0,0,0,2,0],
		[0,0,1,0,0,0,0,2,0,0],
		[0,0,0,1,0,0,2,0,0,0],
		[0,0,0,0,1,2,0,0,0,0],
		[0,0,0,0,2,1,0,0,0,0],
		[0,0,0,2,0,0,1,0,0,0],
		[0,0,2,0,0,0,0,1,0,0],
		[0,2,0,0,0,0,0,0,1,0],
		[2,0,0,0,0,0,0,0,0,1],
	],
\end{lstlisting}

After choosing the number of packets per matrix position, the next field specifies the colors of each entry in the matrix to grey (0), blue (1), or red (2). These colors can be an important aid for illustrating important cybersecurity concepts such as internal networks (blue) and adversarial networks (red).

\begin{lstlisting}[basicstyle=\ttfamily\scriptsize]
	"traffic_matrix_colors":[
		[0,0,0,0,0,0,2,2,2,2],
		[0,0,0,0,0,0,2,2,2,2],
		[0,0,0,0,0,0,2,2,2,2],
		[0,0,0,0,0,0,2,2,2,2],
		[0,0,0,0,0,0,0,0,0,0],
		[0,0,0,0,0,0,0,0,0,0],
		[1,1,1,1,0,0,0,0,0,0],
		[1,1,1,1,0,0,0,0,0,0],
		[1,1,1,1,0,0,0,0,0,0],
		[1,1,1,1,0,0,0,0,0,0],
	],
\end{lstlisting}

The final fields in the JSON specify if there is a question, what the question is, the possible answers, and corresponding correct answer.  \emph{Traffic Warehouse} will randomize the list that has the answers when they are displayed, so the first element will not always be the first option given.  Our choice to have three available multiple choice answers was deliberate as there is some evidence indicating that using a three choice multiple choice questions is beneficial in balancing the quality multiple choice questions against devaluing the assessment of the student's knowledge \cite{doi:10.1177/0013164493053004013} \cite{doi:10.1177/0013164493053003021} \cite{DEHNAD2014398}. The ability to toggle a question on and off allows for a more interactive experience where an educator can have an open discussion or prompt an entire class through online polls.

\begin{lstlisting}[basicstyle=\ttfamily\scriptsize]
	"has_question":true,
	"question":"How many packets did WS1 send to ADV4?",
	"answers":["0", "1", "2",],
	"correct_answer_element":2,
\end{lstlisting}

JSON is a common, lightweight, human readable data interchange format that is well known in the information technology and cybersecurity fields.  JSON is a plaintext file so the template can be edited with a simple text editor. Likewise, being plaintext allows for any security review to be accomplished quickly and efficiently. If a learning module needs to be brought into a restricted area it can be easily done so on printed paper and reviewed and approved by the appropriate security team member, then simply hand typed back on to the restricted machine.

\section{Game Technology}

A variety of technologies can be used to create a video game environment.   Game engines allow a programmer to code the environment, presentation, and dynamics of the game (see Table~\ref{tab:godot-unity-unreal-compare}).  The 3D visual objects or \emph{assets} in the game are typically created with a modeling program (see Table~\ref{tab:blender-maya-mv-compare}).  Care must be used to select  game engine and modeling technologies  that are appropriate to goals and scale of the game development effort.  In this case, for a simple educational game, the emphasis is on availability and ease-of-use so that others can readily build on the work.

\begin{table}[]
\resizebox{\columnwidth}{!}{%
\begin{tabular}{c|c|c|c|}
\cline{2-4}
\multicolumn{1}{l|}{}                                                               & Godot                                                                                        & Unity                                                                                                   & Unreal                                                                                               \\ \hline
\multicolumn{1}{|c|}{Cost}                                                          & \cellcolor[HTML]{67FD9A}Always Free                                                          & \cellcolor[HTML]{67FD9A}\begin{tabular}[c]{@{}c@{}}Free when making \\ less than \$100k/yr\end{tabular} & \cellcolor[HTML]{67FD9A}\begin{tabular}[c]{@{}c@{}}Free when making \\ less than \$1mil\end{tabular} \\ \hline
\multicolumn{1}{|c|}{Language Used}                                                 & \cellcolor[HTML]{67FD9A}C\#, GDScript                                                        & \cellcolor[HTML]{FFFE65}C\#                                                                             & \cellcolor[HTML]{FFFE65}C++                                                                          \\ \hline
\multicolumn{1}{|c|}{Can Import .obj}                                               & \cellcolor[HTML]{67FD9A}Yes                                                                  & \cellcolor[HTML]{67FD9A}Yes                                                                             & \cellcolor[HTML]{67FD9A}Yes                                                                          \\ \hline
\multicolumn{1}{|c|}{\begin{tabular}[c]{@{}c@{}}Exports to\\ Platform\end{tabular}} & \cellcolor[HTML]{67FD9A}\begin{tabular}[c]{@{}c@{}}HTML5, Windows, \\ Mac, *NIX\end{tabular} & \cellcolor[HTML]{67FD9A}\begin{tabular}[c]{@{}c@{}}HTML5, Windows, \\ Mac, *NIX\end{tabular}            & \cellcolor[HTML]{67FD9A}\begin{tabular}[c]{@{}c@{}}HTML5, Windows, \\ Mac, *NIX\end{tabular}         \\ \hline
\multicolumn{1}{|c|}{Online Tutorials}                                              & \cellcolor[HTML]{FFFE65}Some                                                                 & \cellcolor[HTML]{67FD9A}Many                                                                            & \cellcolor[HTML]{67FD9A}Many                                                                         \\ \hline
\multicolumn{1}{|c|}{Asset Store}                                                   & \cellcolor[HTML]{FFFE65}Almost non-existent                                                  & \cellcolor[HTML]{67FD9A}Many high quality assets                                                        & \cellcolor[HTML]{67FD9A}Many high quality assets                                                     \\ \hline
\end{tabular}%
}
\caption{Comparison between the Godot engine and two other industry standards, Unity and Unreal. We compare the cost, the language used in-engine, the ability to import .obj files, what platforms that can the engine build for, the availability of online tutorials and training courses, and how robust the asset store is for purchasing premade assets.}
\label{tab:godot-unity-unreal-compare}
\end{table}

\begin{table}[]
\resizebox{\columnwidth}{!}{%
\begin{tabular}{c|c|c|c|}
\cline{2-4}
\multicolumn{1}{l|}{}                                                               & MagicaVoxel                                                                                             & Blender                                                                                            & Maya                                                                                               \\ \hline
\multicolumn{1}{|c|}{Cost}                                                          & \cellcolor[HTML]{67FD9A}Free to use                                                                     & \cellcolor[HTML]{67FD9A}Free to use                                                                & \cellcolor[HTML]{FD6864}\$1,875/yr                                                                 \\ \hline
\multicolumn{1}{|c|}{\begin{tabular}[c]{@{}c@{}}Model \\ Creation\end{tabular}}     & \cellcolor[HTML]{67FD9A}\begin{tabular}[c]{@{}c@{}}LEGO-like \\ voxel building\end{tabular}             & \cellcolor[HTML]{FCFF2F}\begin{tabular}[c]{@{}c@{}}Polygon mesh, \\ digital sculpting\end{tabular} & \cellcolor[HTML]{FCFF2F}\begin{tabular}[c]{@{}c@{}}Polygon mesh, \\ digital sculpting\end{tabular} \\ \hline
\multicolumn{1}{|c|}{\begin{tabular}[c]{@{}c@{}}Texture \\ Creation\end{tabular}}   & \cellcolor[HTML]{67FD9A}\begin{tabular}[c]{@{}c@{}}Paint-by-voxel,  \\ place colored voxel\end{tabular} & \cellcolor[HTML]{FCFF2F}\begin{tabular}[c]{@{}c@{}}UV Unwrapping, \\ paint-on-model\end{tabular}   & \cellcolor[HTML]{FCFF2F}\begin{tabular}[c]{@{}c@{}}UV Unwrapping, \\ paint-on-model\end{tabular}   \\ \hline
\multicolumn{1}{|c|}{Animation}                                                     & \cellcolor[HTML]{67FD9A}\begin{tabular}[c]{@{}c@{}}Simple \\ animations\end{tabular}                    & \cellcolor[HTML]{67FD9A}\begin{tabular}[c]{@{}c@{}}Advanced \\ animations\end{tabular}             & \cellcolor[HTML]{67FD9A}\begin{tabular}[c]{@{}c@{}}Advanced \\ animations\end{tabular}             \\ \hline
\multicolumn{1}{|c|}{\begin{tabular}[c]{@{}c@{}}Can export \\ to .obj\end{tabular}} & \cellcolor[HTML]{67FD9A}Yes                                                                             & \cellcolor[HTML]{67FD9A}Yes                                                                        & \cellcolor[HTML]{67FD9A}Yes                                                                        \\ \hline
\end{tabular}%
}
\caption{Comparison between two industry standard 3D modeling programs and MagicaVoxel. We compare the cost, how models are made, how textures are created, animation capability, and if they can export to the model format we need for our game engine.}
\label{tab:blender-maya-mv-compare}
\end{table}

\begin{figure}
	\centering
	\begin{subfigure}{1\columnwidth}
		\lstset{language=[Sharp]C}
		\begin{lstlisting} [basicstyle=\ttfamily\scriptsize]
		static void HelloWorld() 
		{
  			Console.WriteLine("Hello, world!");
		}

		static void Main(string[] args)
		{
  			MyMethod();
		}
		\end{lstlisting}

		\caption{C\#}
	\end{subfigure}%
	\\
	~
	\\
	\begin{subfigure}{1\columnwidth}
		\lstset{language=Python}
		\begin{lstlisting}[basicstyle=\ttfamily\scriptsize]
		def HelloWorld():
			print('Hello, world!')
			
		HelloWorld()
		\end{lstlisting}
		\caption{Python}
	\end{subfigure}%
	\\
	~
	\\
	\begin{subfigure}{1\columnwidth}
		\lstset{language=Python}
		\begin{lstlisting}[basicstyle=\ttfamily\scriptsize]
		func _ready():
			HelloWorld()
		
		func HelloWorld():
			print(``Hello, world!'')
		\end{lstlisting}
		\caption{GDScript}
	\end{subfigure}
	\caption{{\bf Hello World in C\#, Python, and GDScript.}  This example illustrates some of the similarities between GDScript and Python when compared to C\#.}
	\label{fig:code-compare}
\end{figure}

\begin{figure}
	\center{\includegraphics[width=0.5\columnwidth]{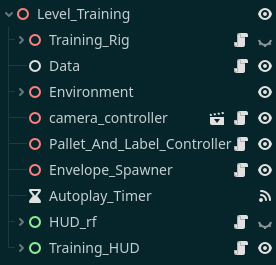}}
	\caption{{\bf Scene Tree Example.}  This shows the scene tree for the training level of \emph{Traffic Warehouse}. It has some elements that are exclusive to the training level but shares many others with a standard level.}
	\label{fig:scenetree}
\end{figure}

\emph{Traffic Warehouse} uses Godot \cite{bradfield2018godot}, a free and open-source game engine.  The engine supports coding in either C\# (for performance) or in a scripting language called GDScript (for ease).  \emph{Traffic Warehouse} is written using GDScript, which is similar to Python and easy to learn (see Figure~\ref{fig:code-compare} for a comparison of C\#, Python, and GDScript).  Godot uses a node and scene based system to build games (Figure~\ref{fig:scenetree}), similar to the way other engines treat prefabs and scriptable objects.  The node/scene system is simple yet powerful, making for a very easy and clean workflow when rapidly developing games.  With the need for  new educational games in the cyberspace community, the ability to rapidly develop new educational games is valuable.  Godot, when compared to staple game engines, such as Unity \cite{goldstone2009unity} or Unreal \cite{lee2016learning}, tends to target lower-end hardware, which makes Godot accessible to larger user base (Table~\ref{tab:godot-unity-unreal-compare}).  Additionally, Godot is designed for fast switching between coding and visual scene editing, further speeding up a developer's workflow.

All of the visual assets for \emph{Traffic Warehouse} were created in MagicaVoxel \cite{MagicaVoxel}, a free-to-use voxel editor which enables rapid development of simple yet appealing 3D assets.  While 3D modeling programs like Maya \cite{ingrassia2008maya} and Blender \cite{flavell2011beginning} are commonly used in industry, they have many excellent features that would not be utilized in a project at this small scale.  When deciding on a tool to create assets our effort focused on a few metrics such as cost, speed of asset creation, and the overall ease of use (see Table~\ref{tab:blender-maya-mv-compare}).  Another benefit to choosing a simpler voxel based modeling program is that future additions can be made by anyone without needing to know how to use a more advanced 3D modeling program.  When provided with a voxel based program, similar canvas size, and a limited color palette a broad audience can create simple game assets in a fairly consistent artistic style.

\section{Implementation}

A key element in educational game design is whether to adopt a direct or metaphorical approach to the learning task \cite{gomez2007using, rieber2008games, law2016puzzle, jackson2016understanding, allegra2021role}.  Network traffic and network traffic matrices are abstract concepts, so a physical metaphor was selected to aid in the understanding of these ideas.   \emph{Traffic Warehouse} presents a stylized shipping warehouse where each entry in the traffic matrix is represented as a grid of shipping pallets on the warehouse floor that can be loaded with boxes (packets) to be shipped.  The shipping warehouse metaphor lends itself to a simple 3D design (floor, pallets, and boxes) and dynamics (placing boxes on pallets).  Like most game engines, Godot can natively read in a JSON file and store it as a dictionary, from which the  \emph{Traffic Warehouse}  learning modules can be instantiated using the shipping warehouse mataphore.

\begin{figure}
	\center{\includegraphics[width=0.5\columnwidth]{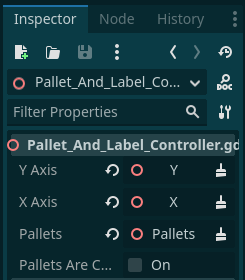}}
	\caption{{\bf Export Variable Example.}  This shows the Inspector tab which allows editing of various properties of our node. By manually exporting several variables they be can edited in this environment.}
	\label{fig:exportvar}
\end{figure}

\begin{figure}
	\center{\includegraphics[width=0.5\columnwidth]{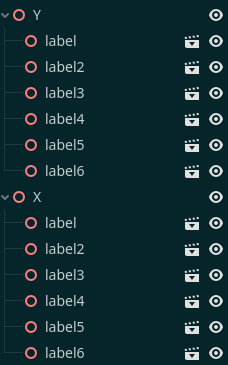}}
	\caption{{\bf X and Y Nodes.}  This shows X and Y nodes and their children, the label nodes.}
	\label{fig:x-y-nodes}
\end{figure}

As an illustrative example of how different aspects of the game were created, consider how the matrix row and column label names and pallet colors are assigned using the JSON learning module file. The following script is broken up into several parts in this paper but is a single file attached to the ``Pallet\_and\_label\_controller'' node in the engine (Figure~\ref{fig:scenetree}).  In Godot a node is the smallest component that can be modified and used to build a scene.  Several export variables are created to allow these variables be dynamically edited without having to edit the script as a whole (Figure~\ref{fig:exportvar}).  Additional variables are set to be assigned as soon as the node is loaded to get the level data and a list of all the child pallets that will be used later.

\lstset{language=Python}
\begin{lstlisting}[basicstyle=\scriptsize\ttfamily]
extends Node3D
@export var y_axis : Node3D
@export var x_axis : Node3D
@export var pallets : Node3D
@export var pallets_are_colored : bool = false
@onready var level_data : Node3D = $"../Data"
@onready var pallet_array : Array = pallets.get_children()
\end{lstlisting}

Next,  an empty array is declared that will hold the pallet color codes. Different materials, or colors, are assigned to different variables.  The different materials are preloaded so they can be quickly assigned to each pallet a set by the traffic matrix colors field from the JSON file.

\begin{lstlisting}[basicstyle=\scriptsize\ttfamily]
var pallet_color_array : Array = []
var pallet_default_material : StandardMaterial3D = preload("res://Assets/Objects/pallet_material.tres")
var pallet_r_material : StandardMaterial3D = preload("res://Assets/Objects/pallet_material_r.tres")
var pallet_b_material : StandardMaterial3D = preload("res://Assets/Objects/pallet_material_b.tres")
var pallet_g_material : StandardMaterial3D = preload("res://Assets/Objects/pallet_material_g.tres")
var pallet_black_material : StandardMaterial3D = preload("res://Assets/Objects/pallet_material_black.tres")
\end{lstlisting}

The next step is to create a ready function.  In Godot, \_ready() is called as soon as the node enters a scene.  It is useful when trying to let the node figure out where it is, to have it do an action immediately or, in this case, gather data in preparation of performing its intended action.  The ready function is pulling the ``traffic\_matrix\_colors'' from the pre-loaded JSON file stored in the ``Data'' node assigned to the variable ``level\_data'' earlier in the script.  After that is complete, set\_labels() is called.

\begin{lstlisting}[basicstyle=\scriptsize\ttfamily]
func _ready():
	for array in level_data.data["traffic_matrix_colors"]:
		pallet_color_array += array
	set_labels()
\end{lstlisting}

The set\_labels() function retrieves the axis names from the ``Data'' node to assign the text labels along both the x and y axes.  Earlier in the script the variables y\_axis and x\_axis were exported, so they could be assigned using the Inspector tab (Figure~\ref{fig:exportvar}).  These variables have been assigned to two nodes, Y and X, that have all of the label nodes under them as children (Figure~\ref{fig:x-y-nodes}).

\begin{lstlisting}[basicstyle=\scriptsize\ttfamily]
func set_labels():
	var y_labels : Array = y_axis.get_children()
	var x_labels : Array = x_axis.get_children()
	if len(y_labels) != len(x_labels):
		printerr("Number of y labels does not match number of x labels!")
	elif len(level_data.data["axis_labels"]) != len(y_labels):
		printerr("Level data does not match number of labels!")
	else:
		var c : int = 0
		for label in level_data.data["axis_labels"]:
			y_labels[c].get_child(1).text = label
			x_labels[c].get_child(1).text = label
			c += 1
\end{lstlisting}

The last function in this example modifies a previously defined boolean controlling whether pallets are in their default or colored states. It is called whenever the toggle pallet color button is clicked. To ensure that each pallet gets the correct color the function iterates over the color array that contains all of the pallets on the level.  This allows a step-by-step change of each pallet color to the one that was assigned in the JSON file.

\begin{lstlisting}[basicstyle=\scriptsize\ttfamily]
func change_pallet_color():
	print("Change pallet color button")
	var c : int = 0
	if pallets_are_colored:
		print("Palets are colored! Making them default")
		for color in pallet_color_array:
			pallet_array[c].get_child(0).material_override = pallet_default_material
			c += 1
		pallets_are_colored = false
	else:
		print("Palets are default! Making them colored")
		for color in pallet_color_array:
			print("Matching color: " + str(color))
			match int(color):
				0: pallet_array[c].get_child(0).material_override = pallet_g_material
				1: pallet_array[c].get_child(0).material_override = pallet_b_material
				2: pallet_array[c].get_child(0).material_override = pallet_r_material
				_: pallet_array[c].get_child(0).material_override = pallet_black_material
			c += 1
		pallets_are_colored = true
\end{lstlisting}

The above example illustrates some aspects of the implementation of \emph{Traffic Warehouse} in the Godot game engine using the JSON learning module file.

\section{Learning Modules}


\begin{figure}
	\centering
	\begin{subfigure}{0.5\columnwidth}
		\center{\includegraphics[width=1.0\columnwidth,trim={0 2cm 10cm 0},clip]{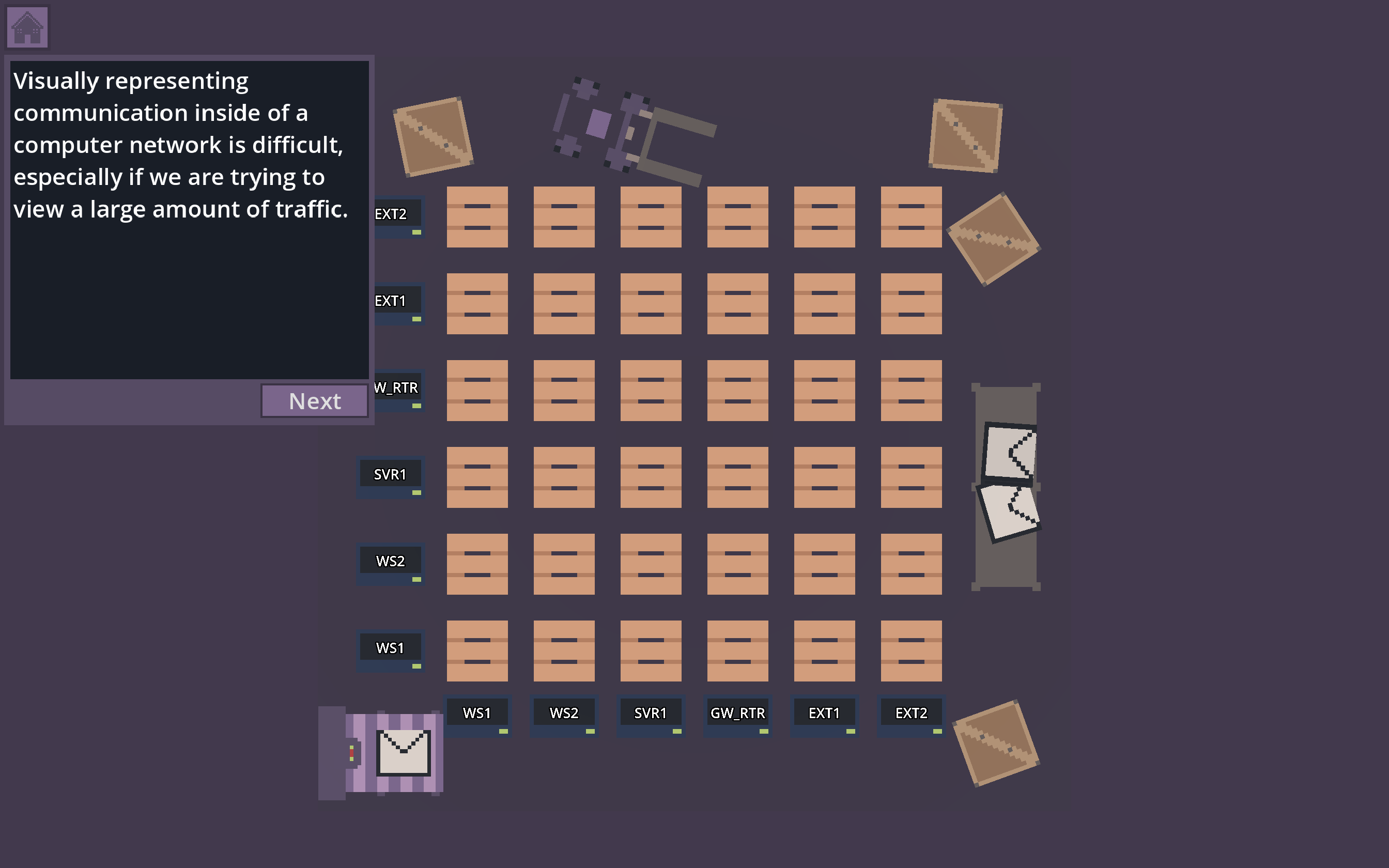}}
		\caption{2D View.}
		\label{fig:mt-2d}
	\end{subfigure}%
	~
	\begin{subfigure}{0.5\columnwidth}
		\center{\includegraphics[width=1.0\columnwidth,trim={0 2cm 10cm 0},clip]{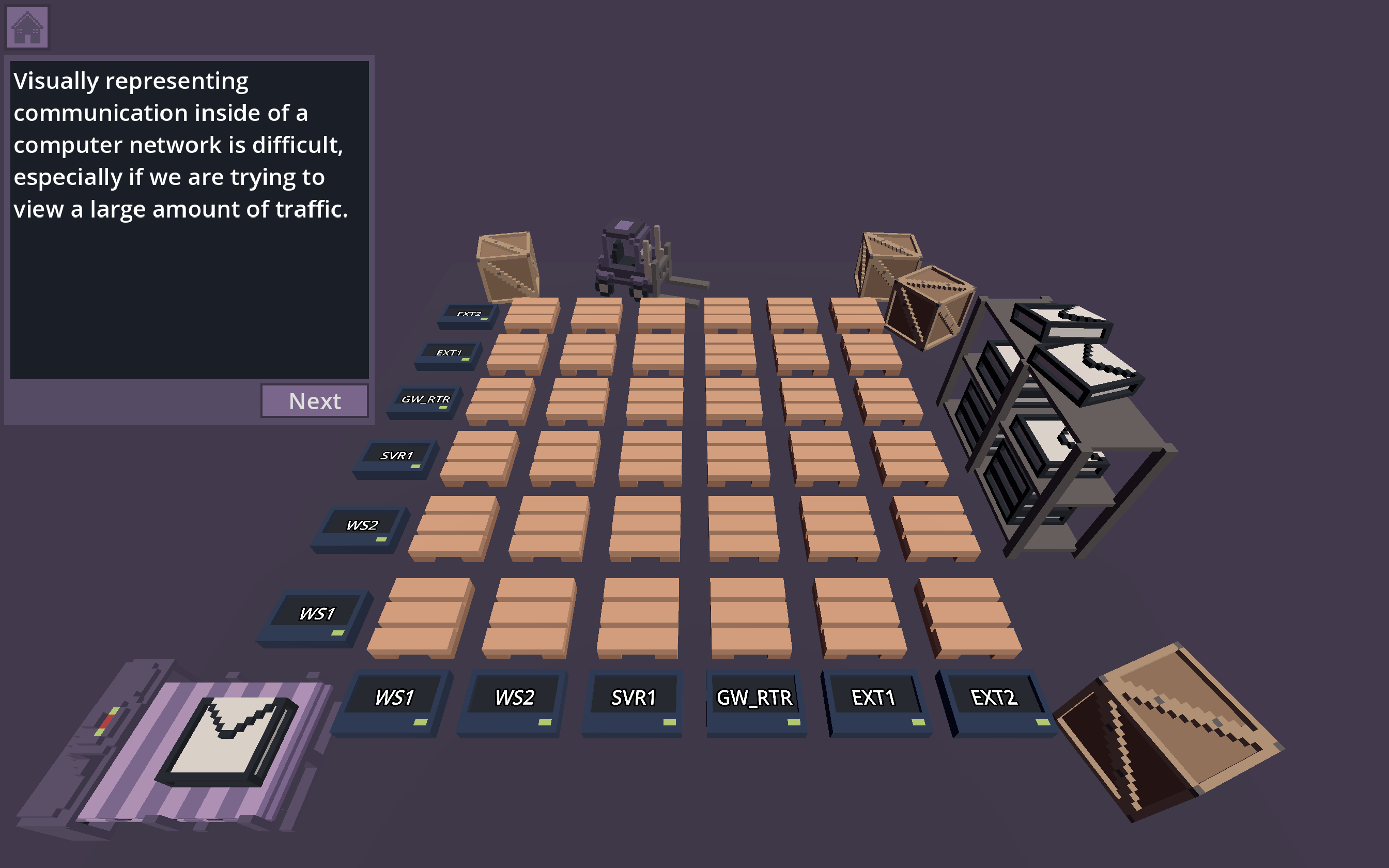}}
		\caption{3D View.}
		\label{fig:mt-3d}
	\end{subfigure}%
	\\
	\begin{subfigure}{0.5\columnwidth}
		\center{\includegraphics[width=1.0\columnwidth,trim={0 2cm 10cm 0},clip]{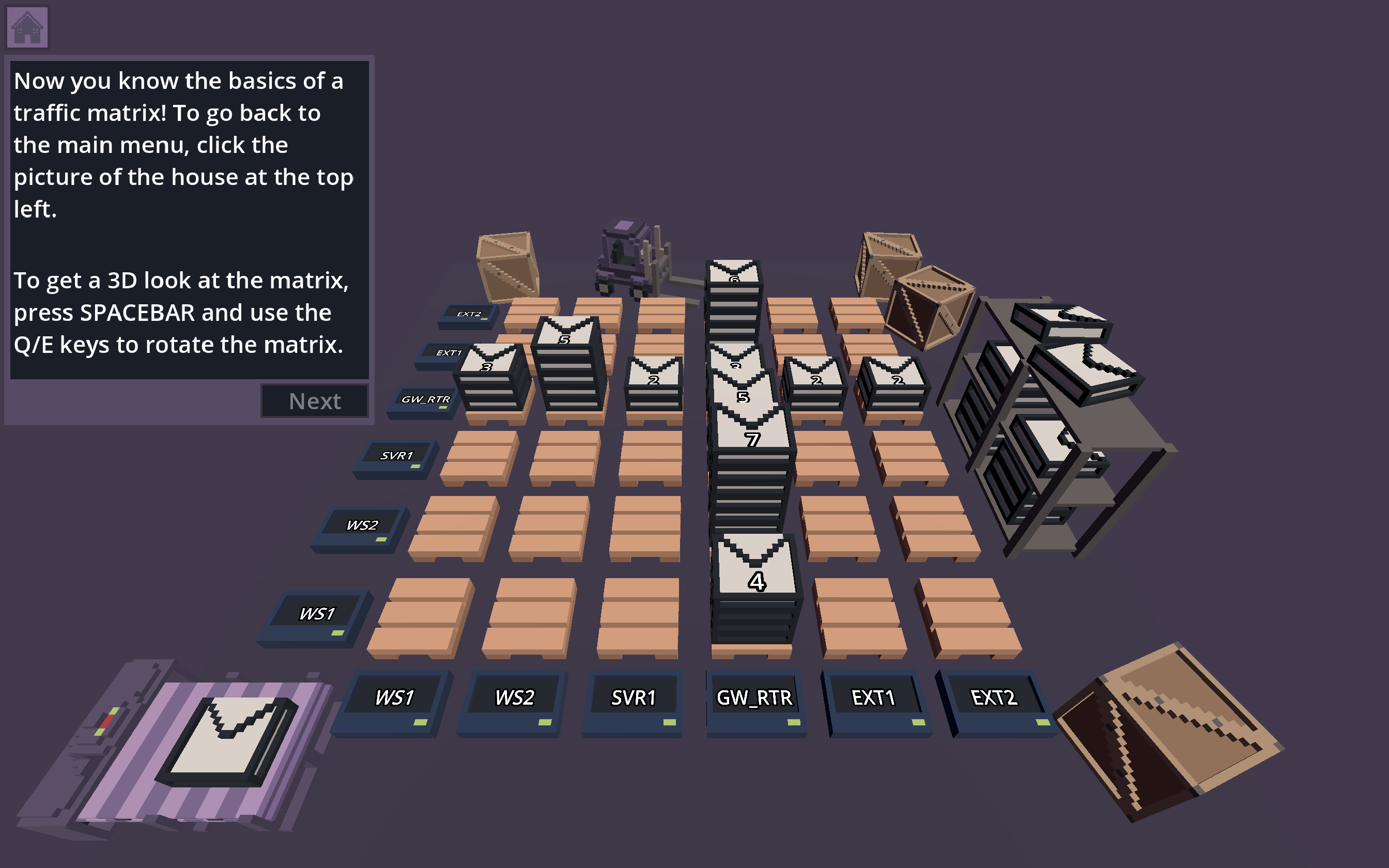}}
		\caption{Packets are all placed.}
		\label{fig:mt-final}
	\end{subfigure}
	\caption{{\bf Traffic Matrix Training.}  Training level of \emph{Traffic Warehouse}. This level walks the player through what a traffic matrix is, why it is valuable to network security personel, and how it will be represented in the game environment.}
	\label{fig:matrix-training}
\end{figure}

\begin{figure}
	\centering
	\begin{subfigure}{0.5\columnwidth}
		\center{\includegraphics[width=1.0\columnwidth,trim={0 5cm 10cm 0},clip]{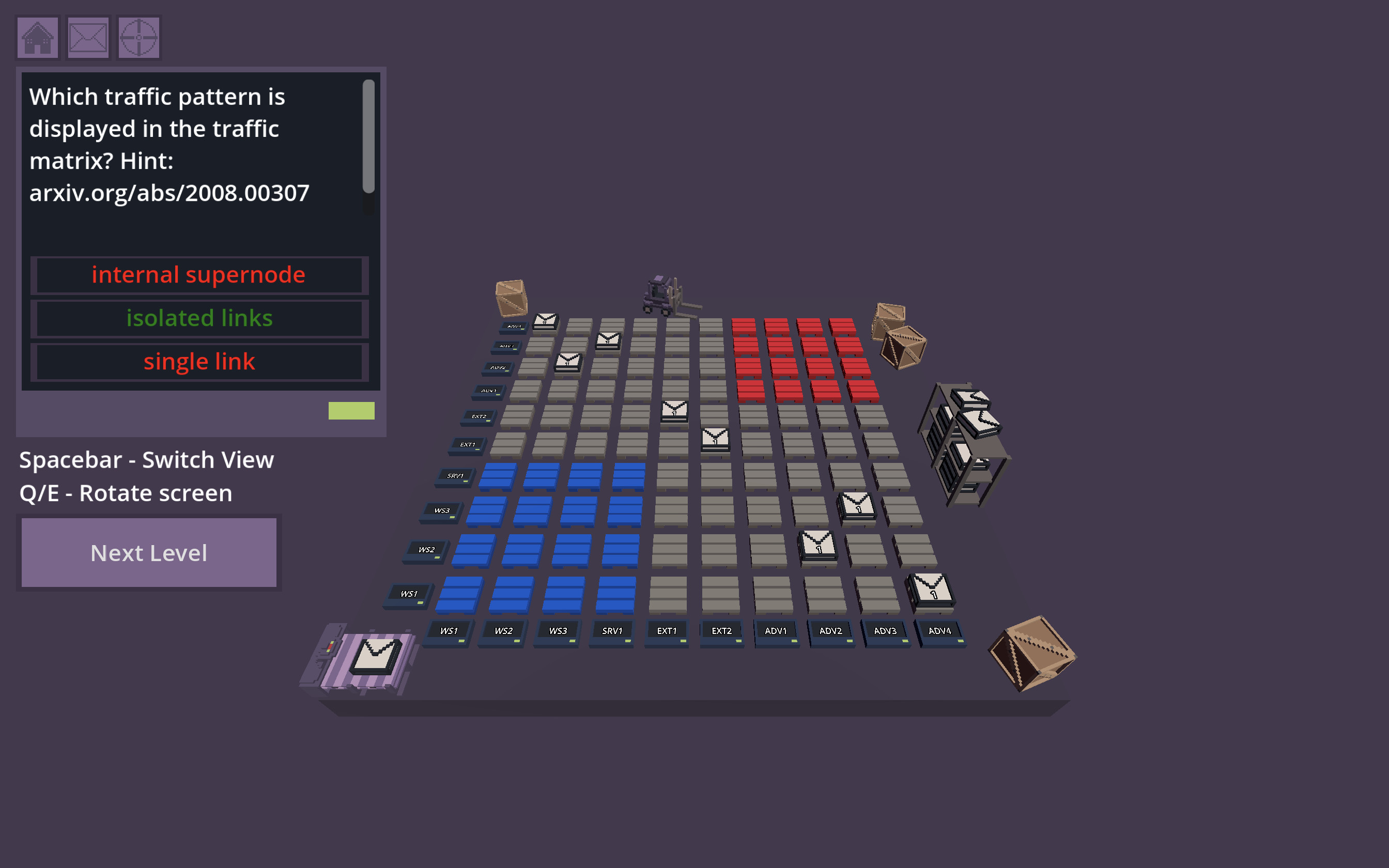}}
		\caption{Isolated Links.}
		\label{fig:tp-isolated-link}
	\end{subfigure}%
	~
	\begin{subfigure}{0.5\columnwidth}
		\center{\includegraphics[width=1.0\columnwidth,trim={0 5cm 10cm 0},clip]{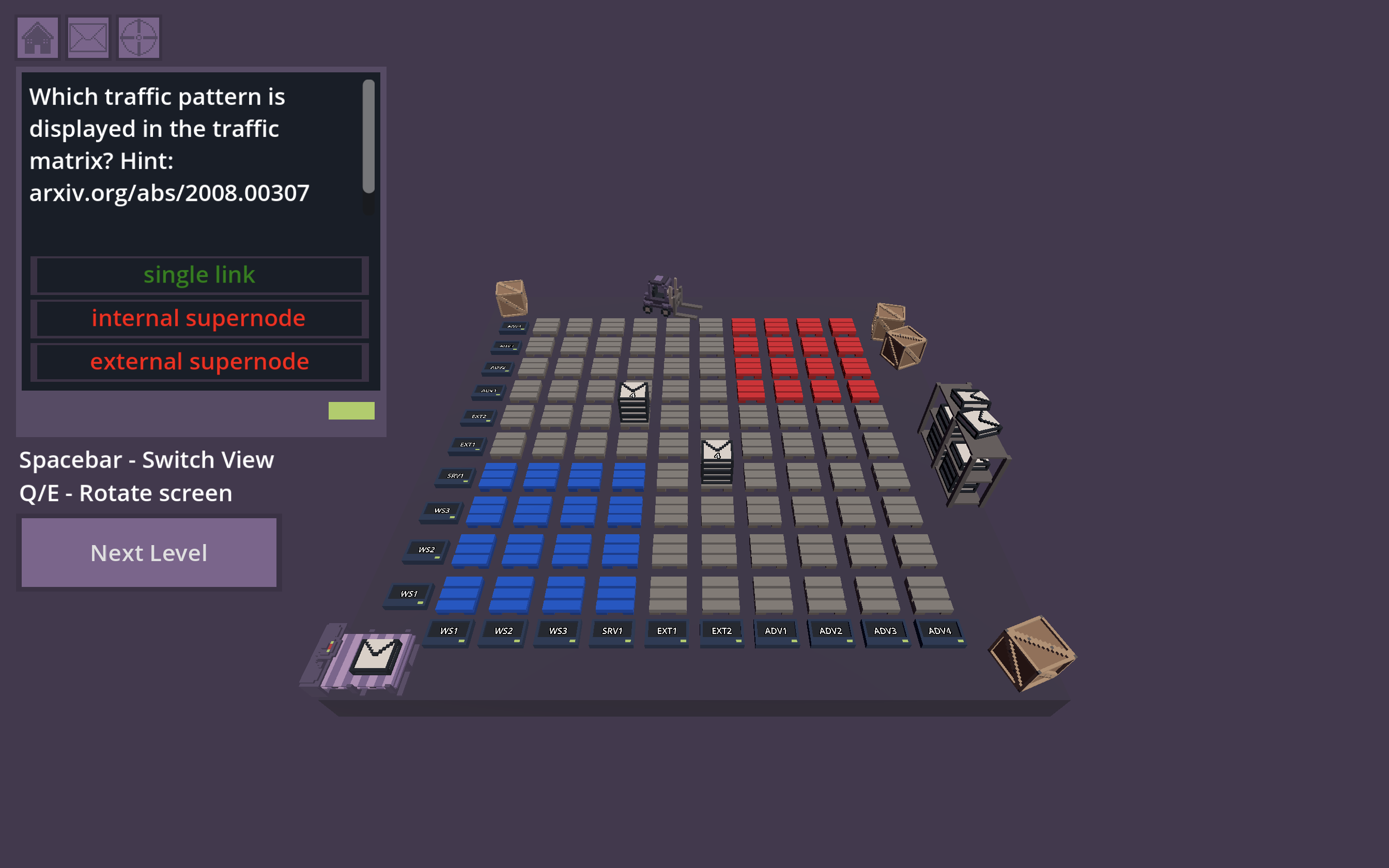}}
		\caption{Single Links.}
		\label{fig:tp-single-link}
	\end{subfigure}%
	\\
	\begin{subfigure}{0.5\columnwidth}
		\center{\includegraphics[width=1.0\columnwidth,trim={0 5cm 10cm 0},clip]{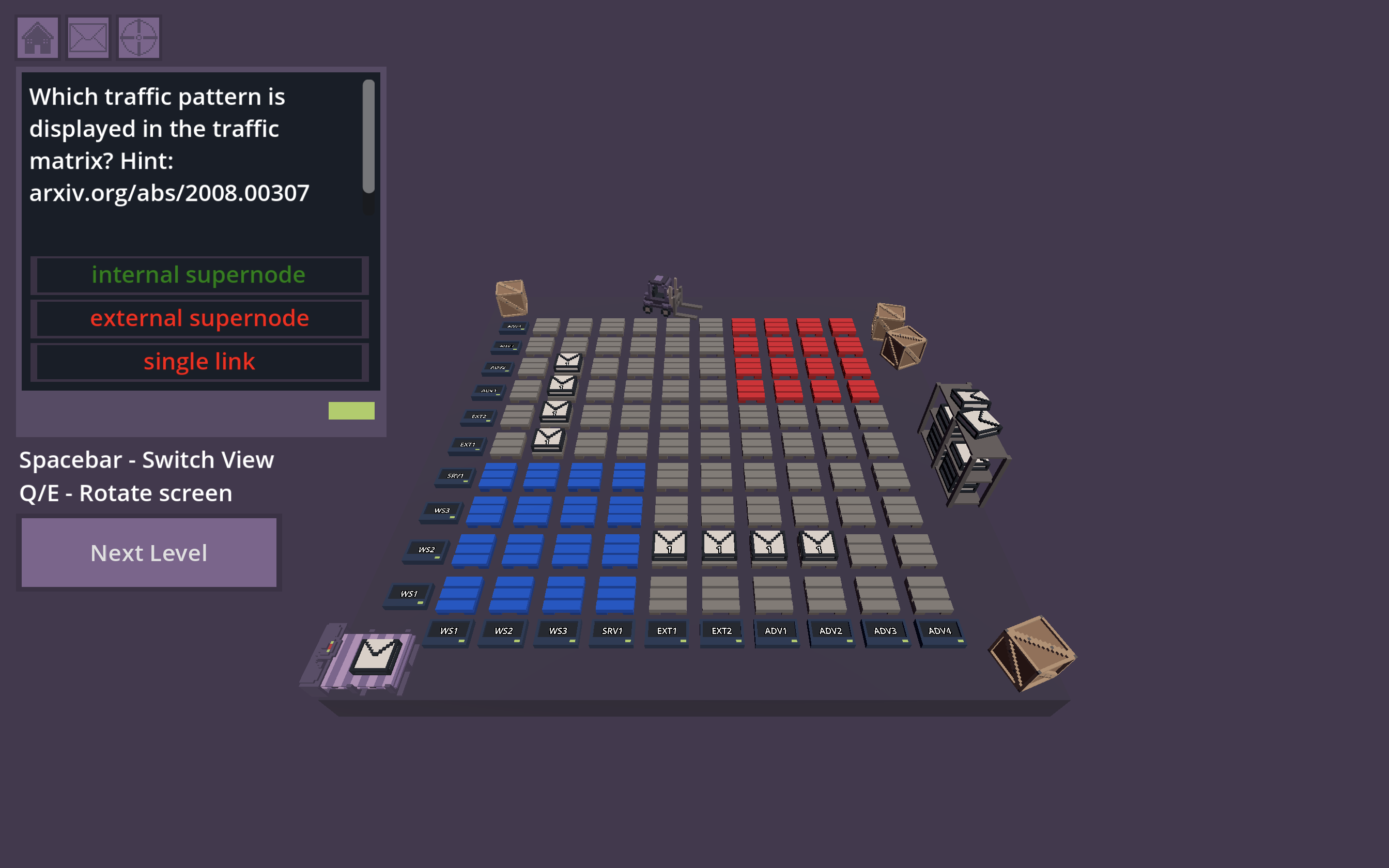}}
		\caption{Internal Supernode.}
		\label{fig:tp-int-supe}
	\end{subfigure}%
	~
	\begin{subfigure}{0.5\columnwidth}
		\center{\includegraphics[width=1.0\columnwidth,trim={0 5cm 10cm 0},clip]{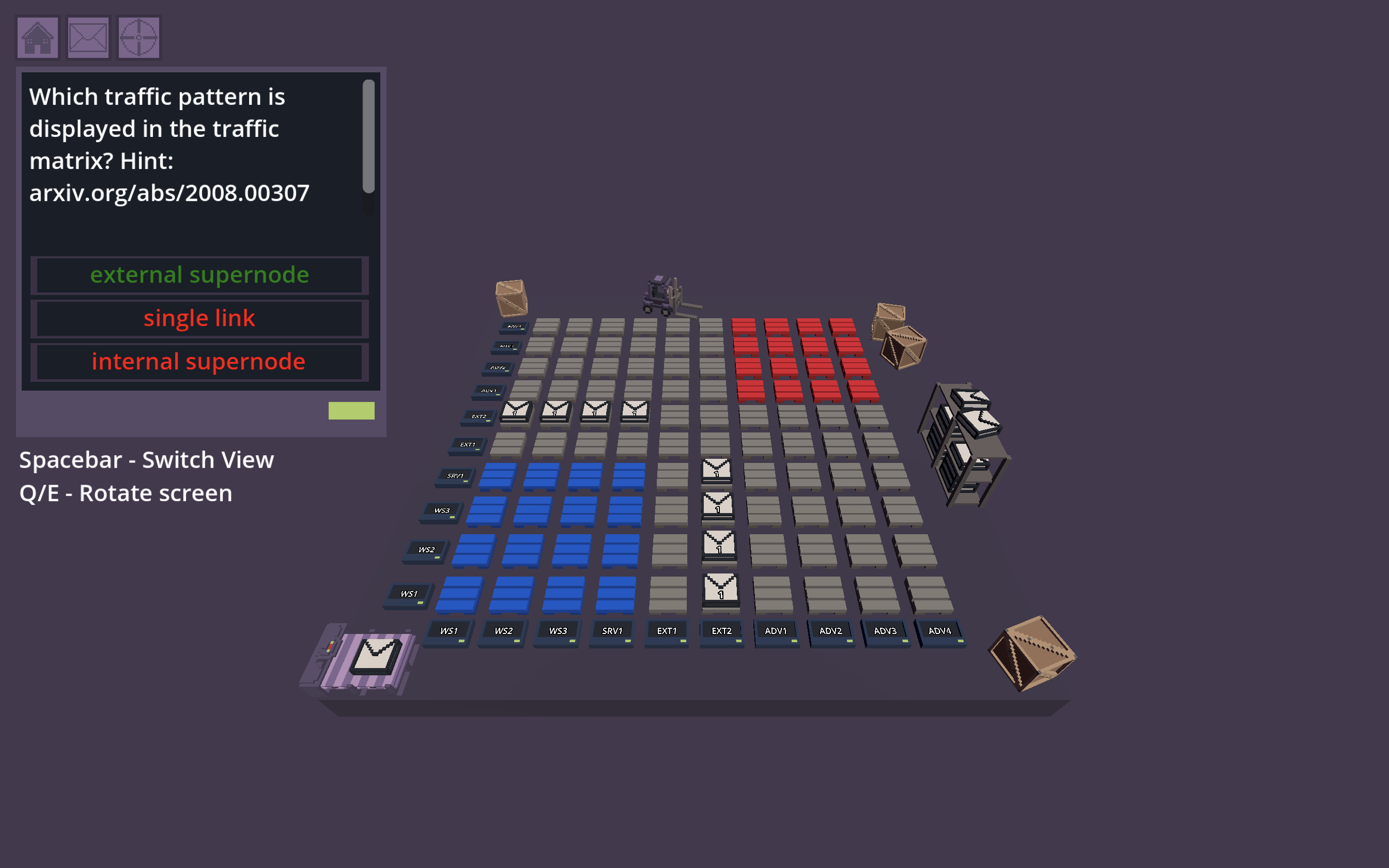}}
		\caption{External Supernode.}
		\label{fig:tp-ext-supe}
	\end{subfigure}
	\caption{{\bf Traffic Topologies.}   Traffic patterns representative of isolated links, single links, internal supernodes, and external supernodes shown on a $10{\times}10$ traffic matrix with a hint to an explanatory reference \cite{Kepner_2020}.}
	\label{fig:traffic-patterns}
\end{figure}

\begin{figure}
	\centering
	\begin{subfigure}{0.5\columnwidth}
		\center{\includegraphics[width=1.0\columnwidth,trim={0 5cm 10cm 0},clip]{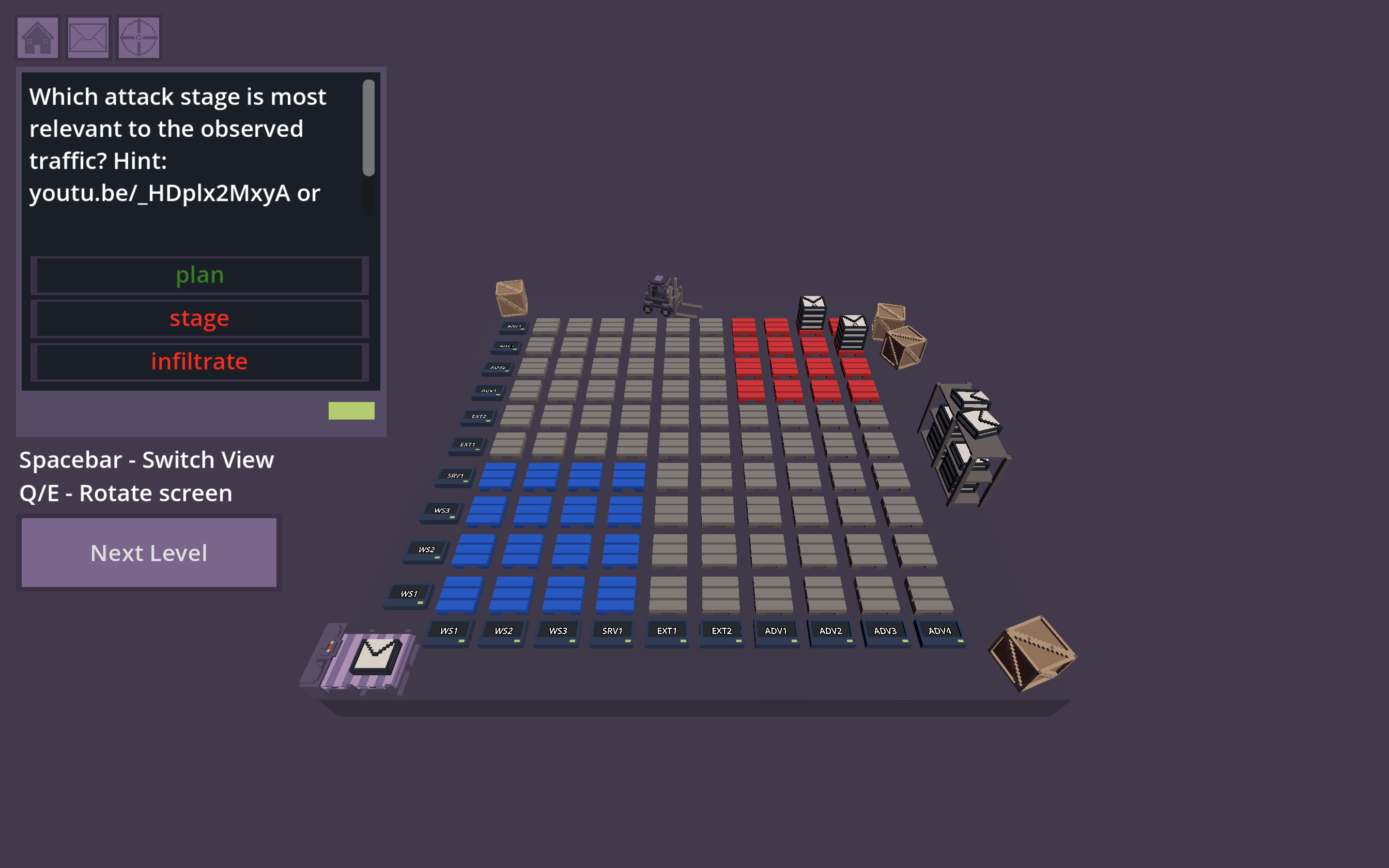}}
		\caption{Planning.}
		\label{fig:na-plan}
	\end{subfigure}%
	~
	\begin{subfigure}{0.5\columnwidth}
		\center{\includegraphics[width=1.0\columnwidth,trim={0 5cm 10cm 0},clip]{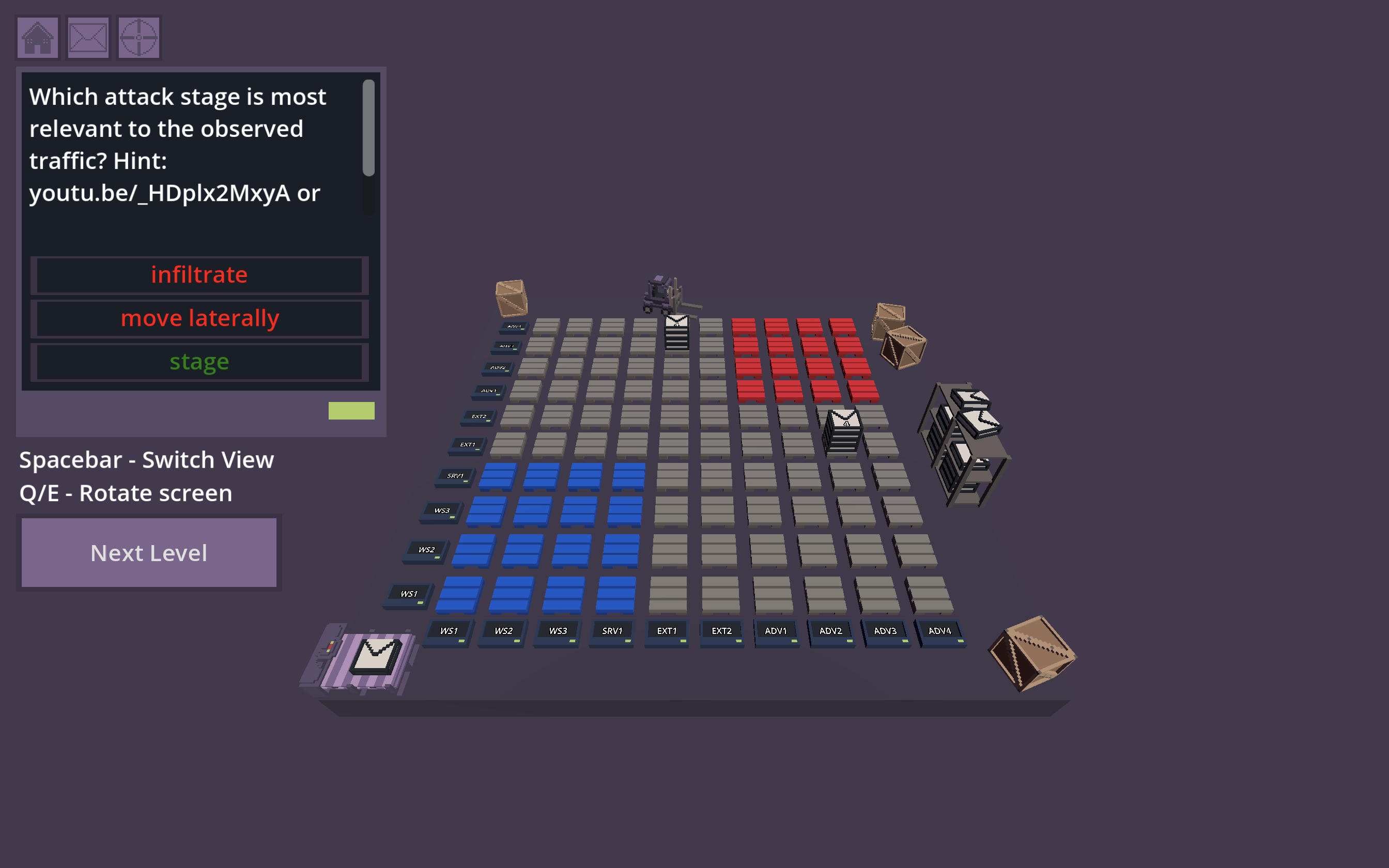}}
		\caption{Staging.}
		\label{fig:na-stage}
	\end{subfigure}%
	\\
	\begin{subfigure}{0.5\columnwidth}
		\center{\includegraphics[width=1.0\columnwidth,trim={0 5cm 10cm 0},clip]{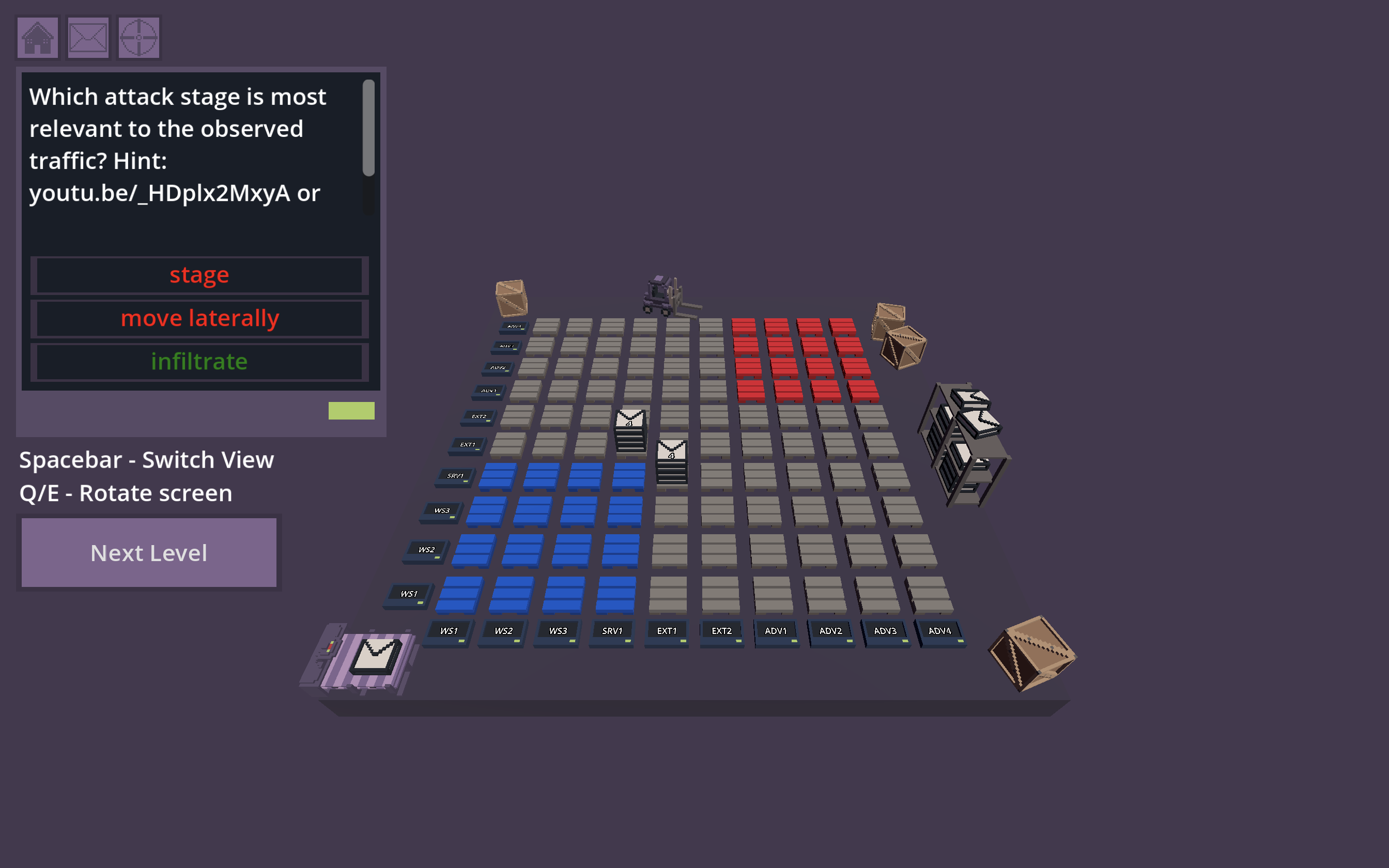}}
		\caption{Infiltration.}
		\label{fig:na-infiltrate}
	\end{subfigure}%
	~
	\begin{subfigure}{0.5\columnwidth}
		\center{\includegraphics[width=1.0\columnwidth,trim={0 5cm 10cm 0},clip]{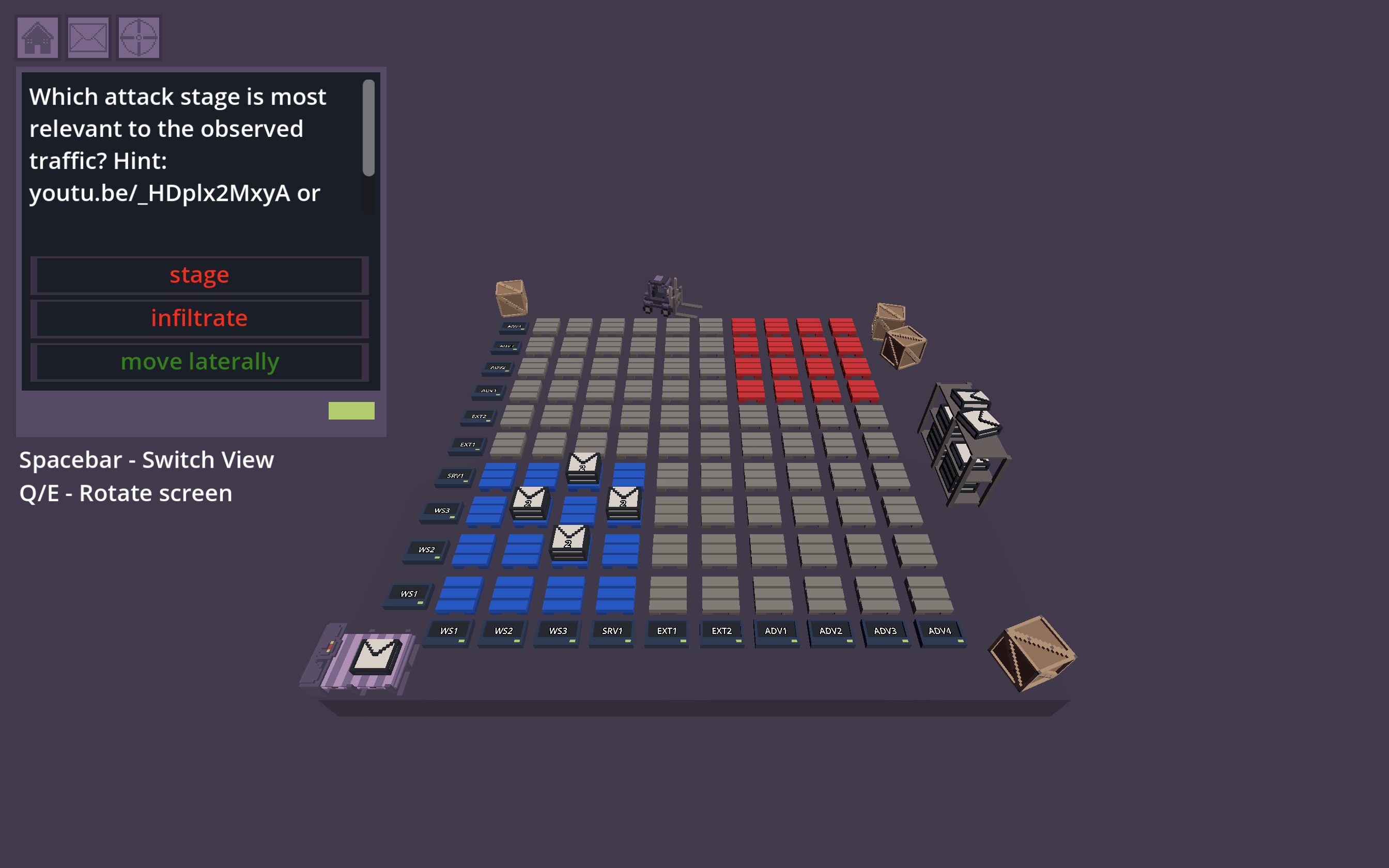}}
		\caption{Lateral Movement.}
		\label{fig:na-lat-move}
	\end{subfigure}
	\caption{{\bf Notional Attack.}  Traffic patterns relevant to the planning, staging, infiltration, and lateral movement stages of an attack shown on a $10{\times}10$ traffic matrix with hints to explanatory references \cite{TEDx_2022,kepner2021zero}}
	\label{fig:notional-attack}
\end{figure}

\begin{figure}
	\centering
	\begin{subfigure}{0.5\columnwidth}
		\center{\includegraphics[width=1.0\columnwidth,trim={0 5cm 10cm 0},clip]{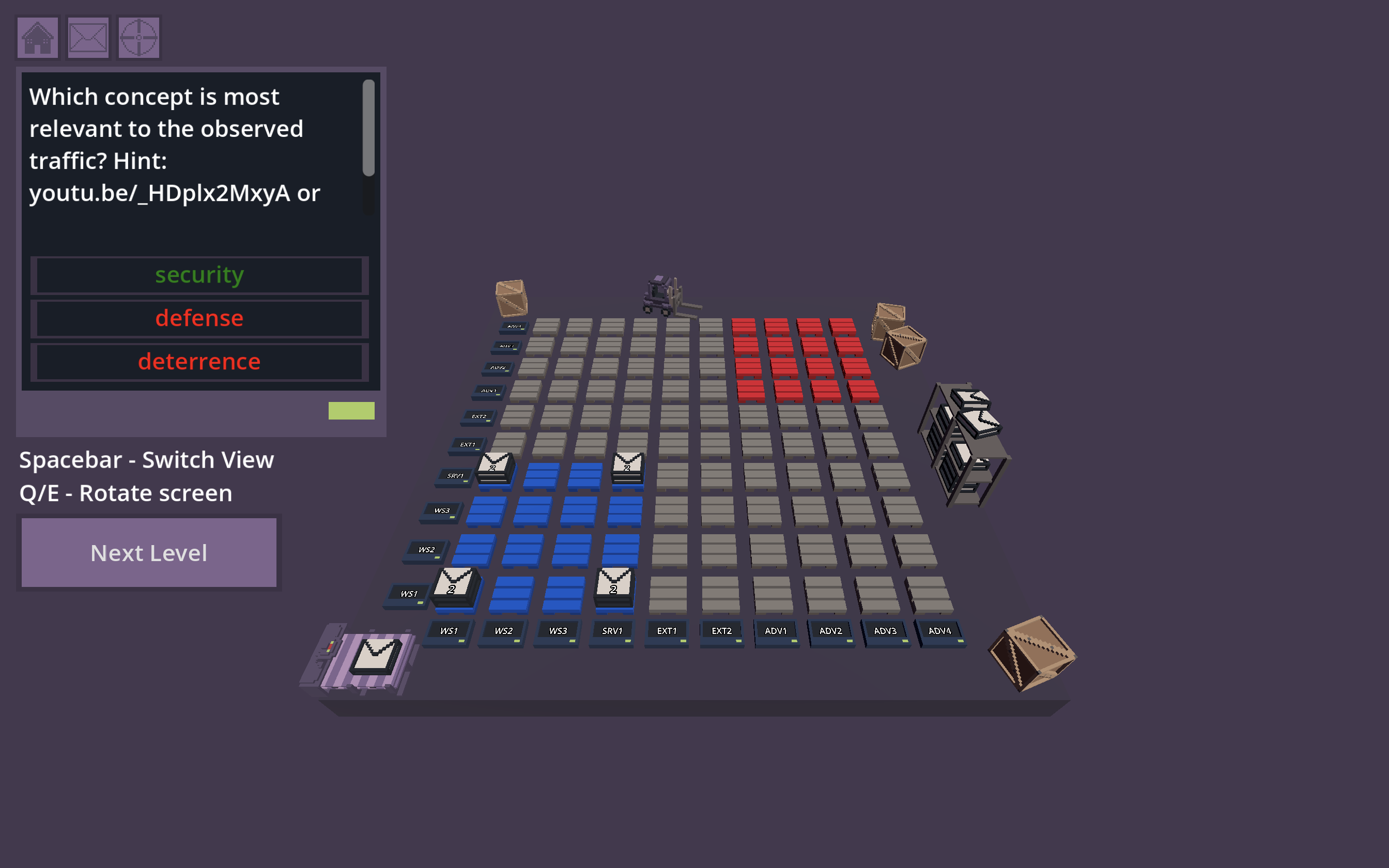}}
		\caption{Security.}
		\label{fig:sdd-sec}
	\end{subfigure}%
	~
	\begin{subfigure}{0.5\columnwidth}
		\center{\includegraphics[width=1.0\columnwidth,trim={0 5cm 10cm 0},clip]{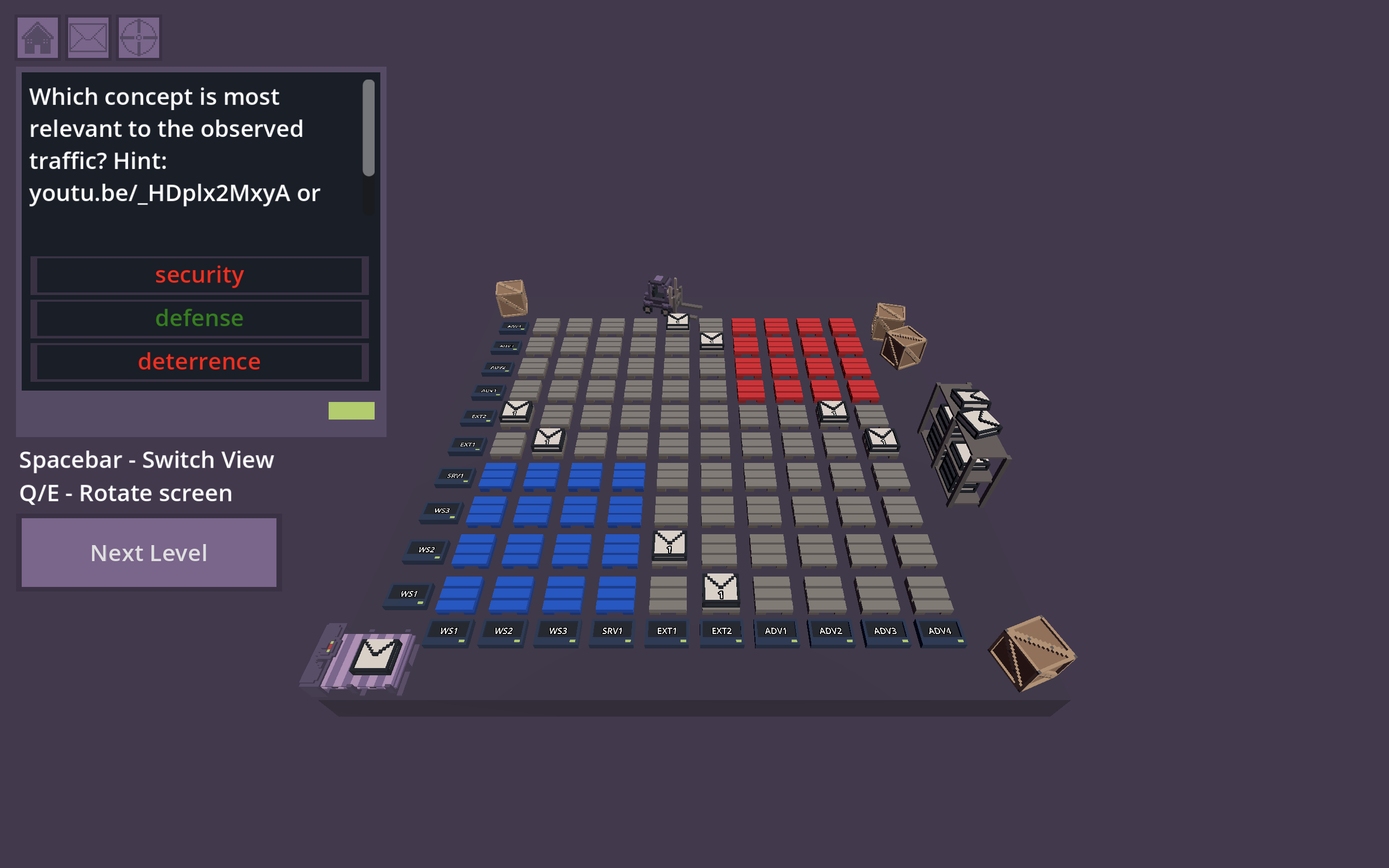}}
		\caption{Defense.}
		\label{fig:sdd-def}
	\end{subfigure}%
	\\
	\begin{subfigure}{0.5\columnwidth}
		\center{\includegraphics[width=1.0\columnwidth,trim={0 5cm 10cm 0},clip]{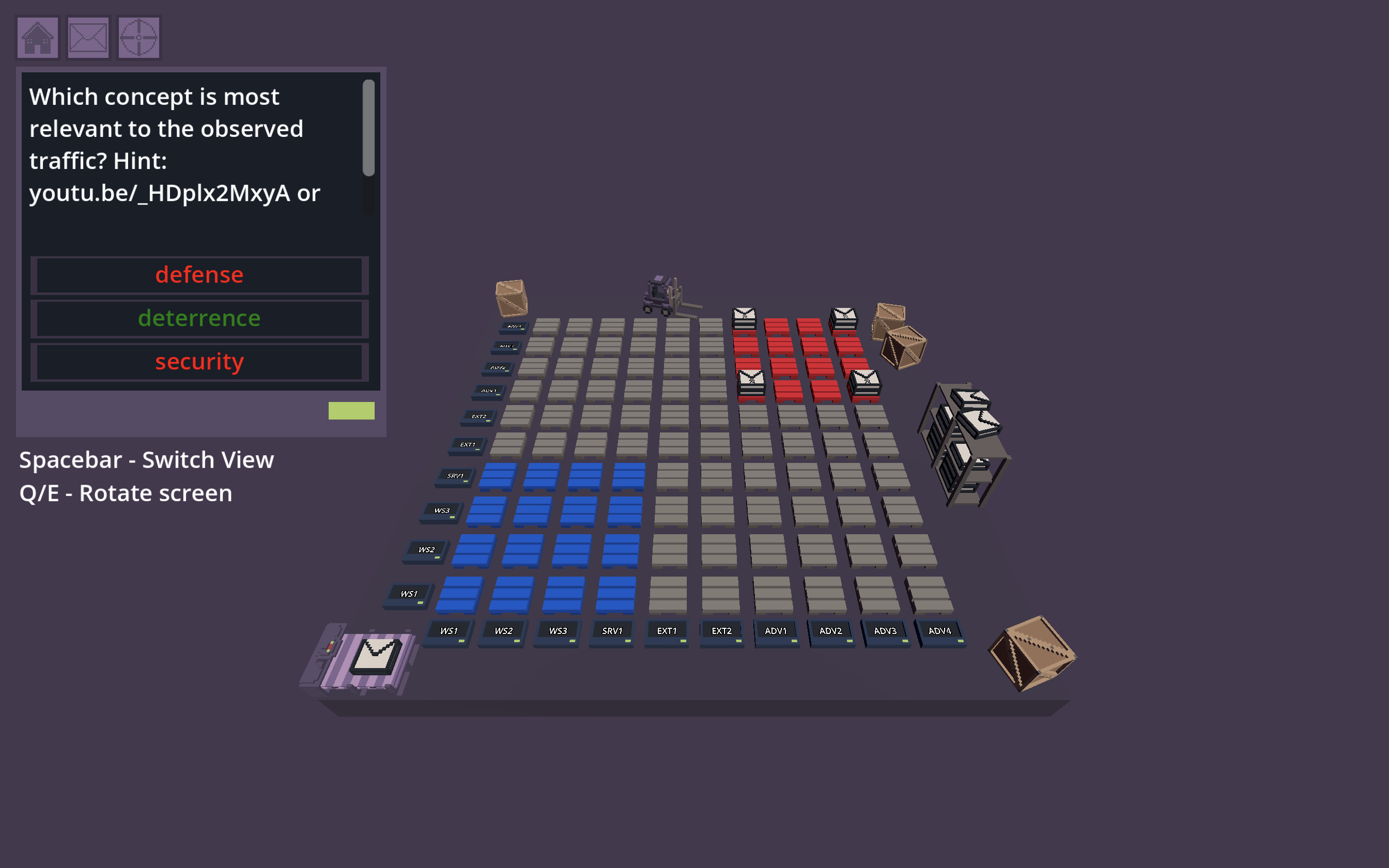}}
		\caption{Deterrence.}
		\label{fig:sdd-deter}
	\end{subfigure}
	\caption{{\bf Network Security, Defense, and Deterrence.}  Traffic patterns relevant to network security, defense, and deterrence approaches shown on a $10{\times}10$ traffic matrix with hints to explanatory references \cite{TEDx_2022,kepner2021zero}}
	\label{fig:sdd}
\end{figure}

\begin{figure}
	\centering
	\begin{subfigure}{0.5\columnwidth}
		\center{\includegraphics[width=1.0\columnwidth,trim={0 5cm 10cm 0},clip]{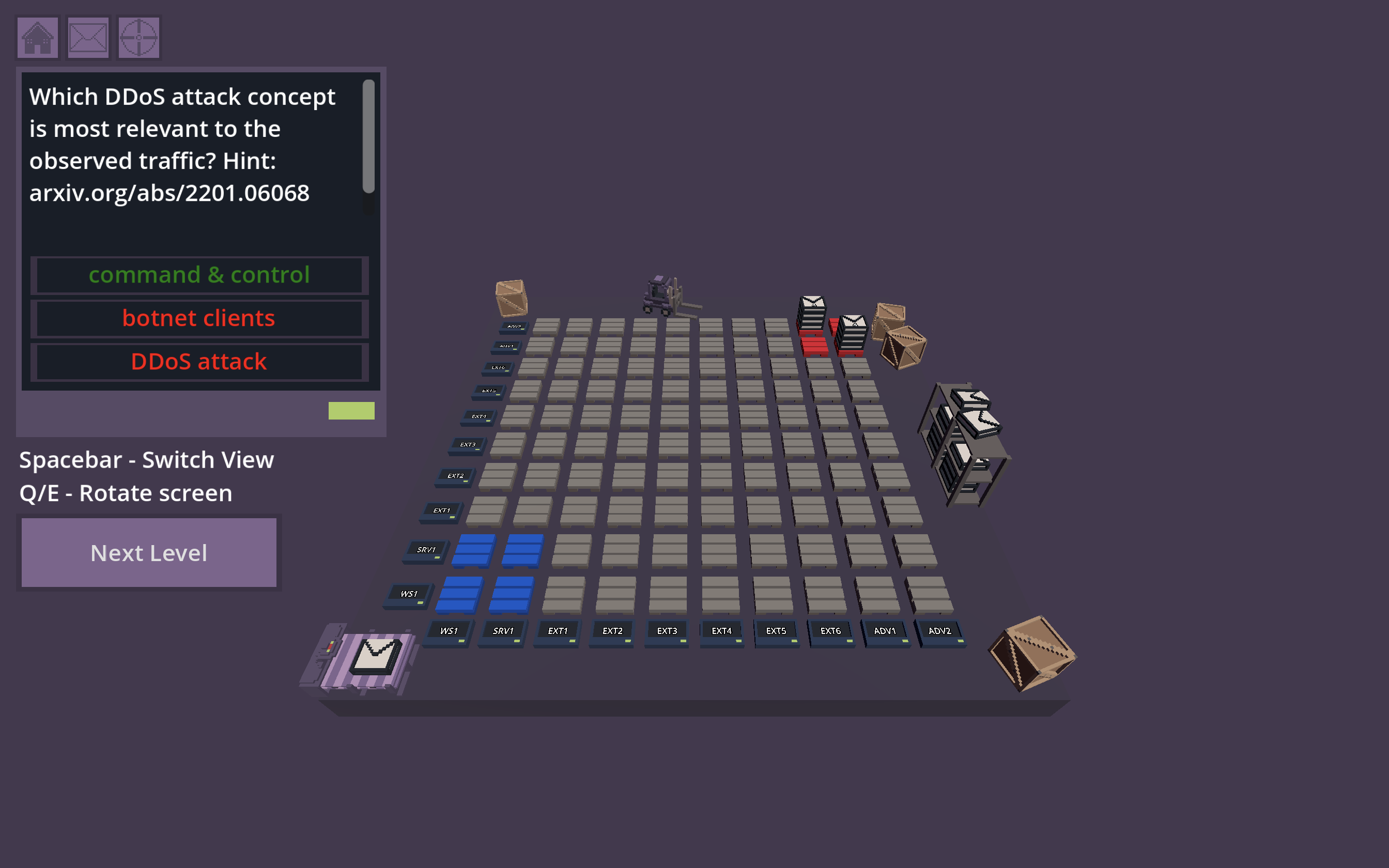}}
		\caption{Command and Control (C2).}
		\label{fig:ddos-c2}
	\end{subfigure}%
	~
	\begin{subfigure}{0.5\columnwidth}
		\center{\includegraphics[width=1.0\columnwidth,trim={0 5cm 10cm 0},clip]{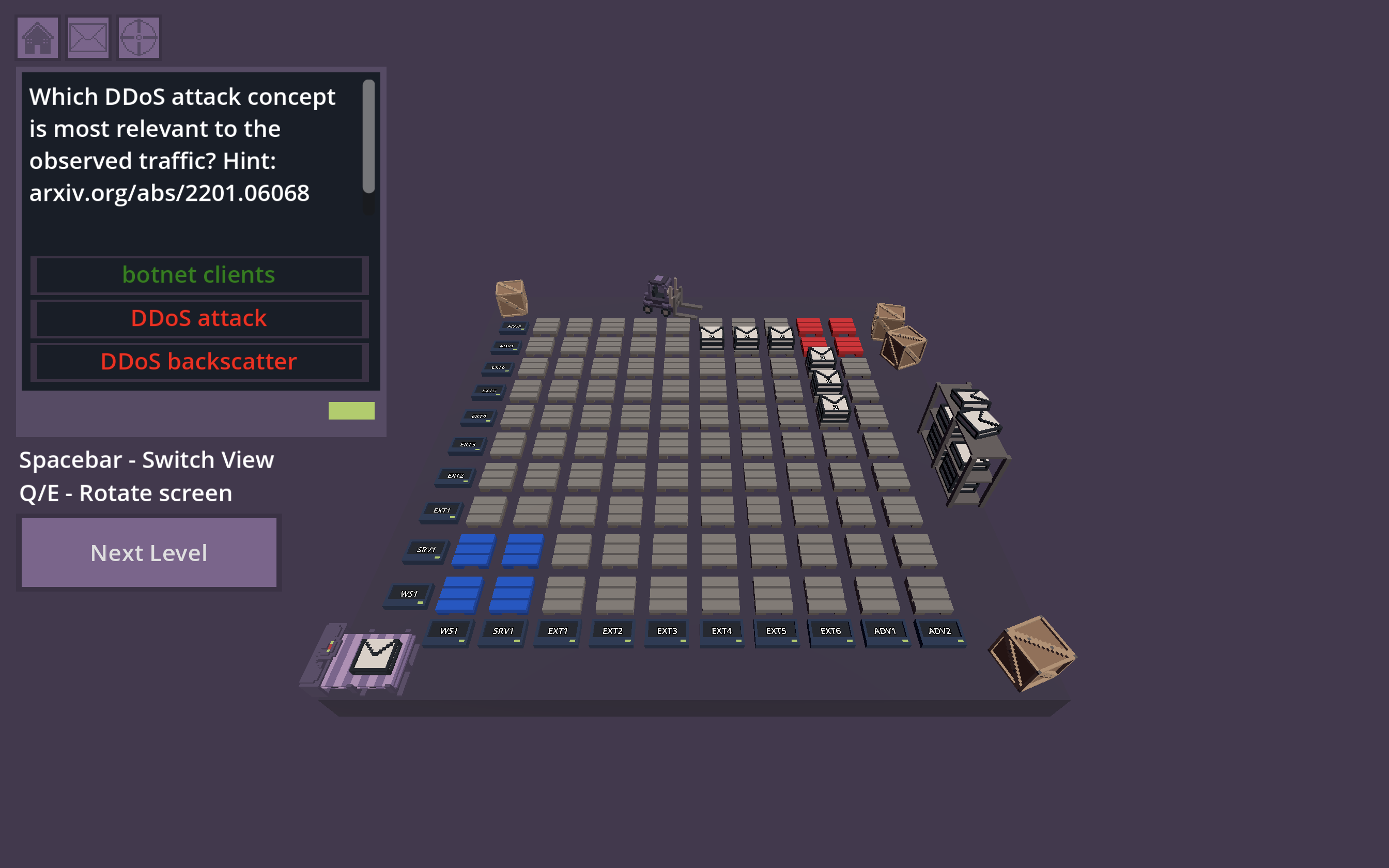}}
		\caption{Botnet Clients.}
		\label{fig:ddos-bot-clients}
	\end{subfigure}%
	\\
	\begin{subfigure}{0.5\columnwidth}
		\center{\includegraphics[width=1.0\columnwidth,trim={0 5cm 10cm 0},clip]{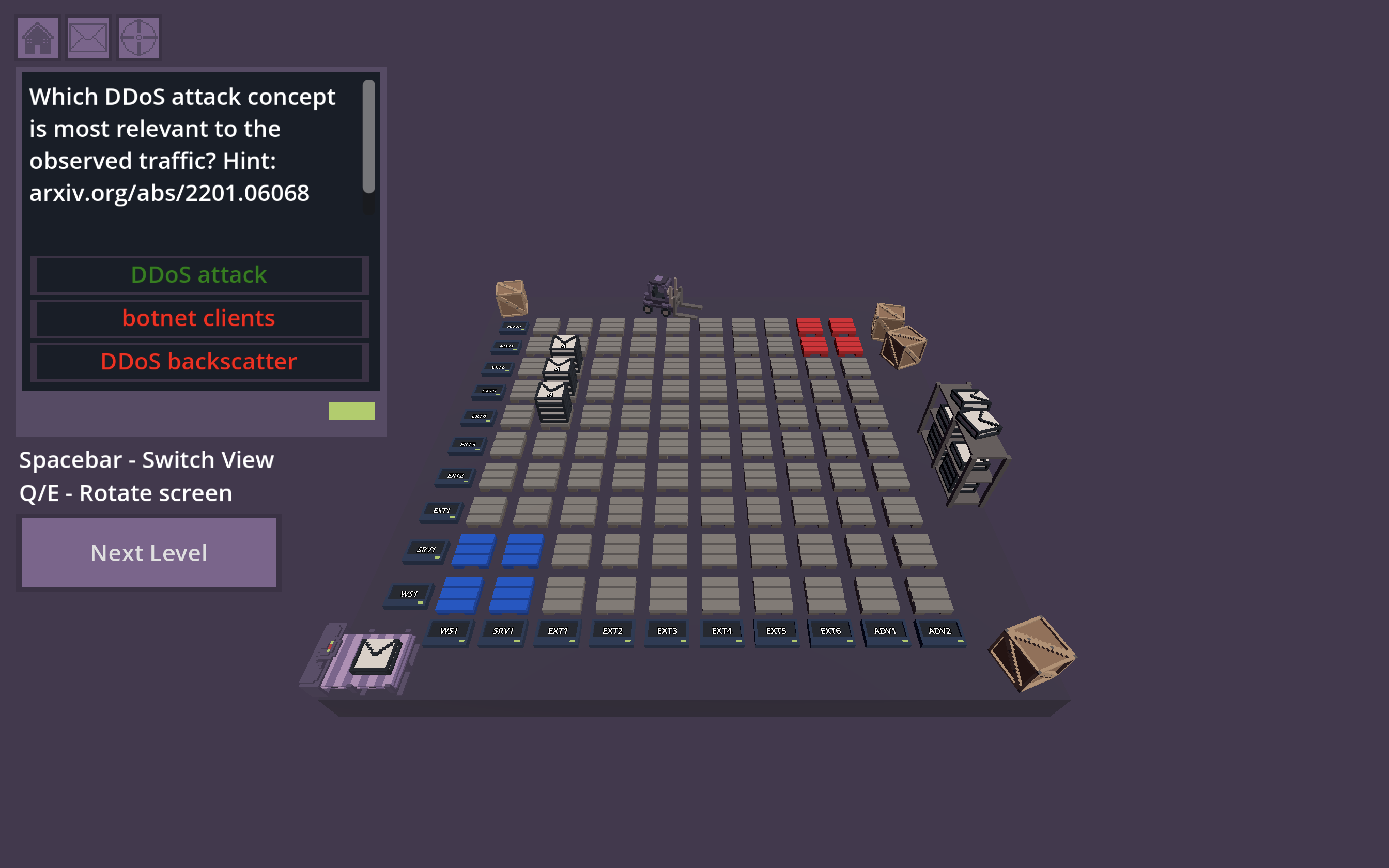}}
		\caption{DDoS Attack.}
		\label{fig:ddos-attack}
	\end{subfigure}%
	~
	\begin{subfigure}{0.5\columnwidth}
		\center{\includegraphics[width=1.0\columnwidth,trim={0 5cm 10cm 0},clip]{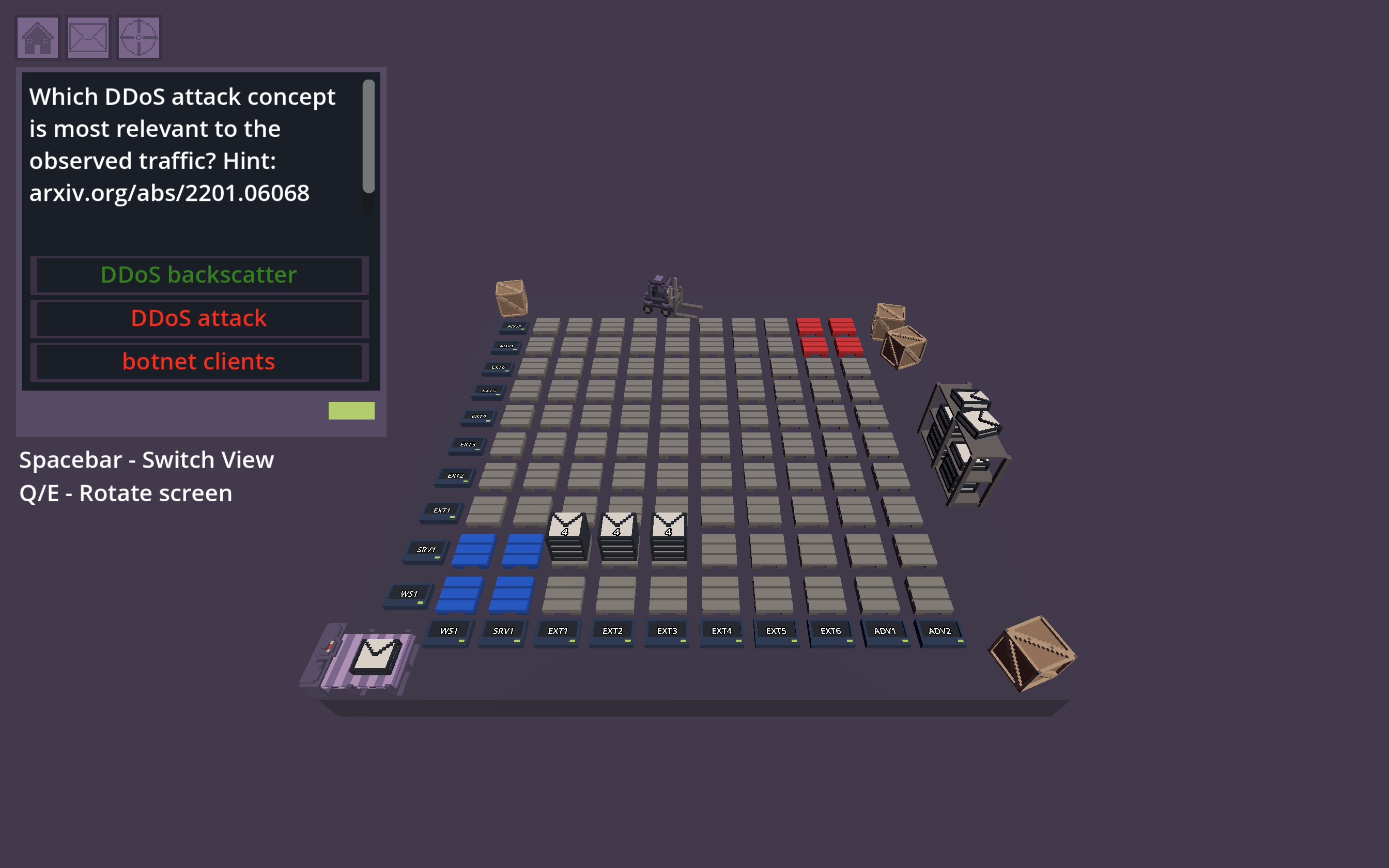}}
		\caption{Backscatter.}
		\label{fig:ddos-backscatter}
	\end{subfigure}
	\caption{{\bf DDoS Attack.}  Traffic patterns relevant to the command-and-control (C2) servers, botnet clients, attack, and backscatter components of a distributed denial-of-service (DDoS) attack shown on a $10{\times}10$ traffic matrix with a hint to an explanatory references \cite{kepner2021zero}}
	\label{fig:ddos}
\end{figure}

\begin{figure}
	\centering
	\begin{subfigure}{0.5\columnwidth}
		\center{\includegraphics[width=1.0\columnwidth,trim={0 5cm 10cm 0},clip]{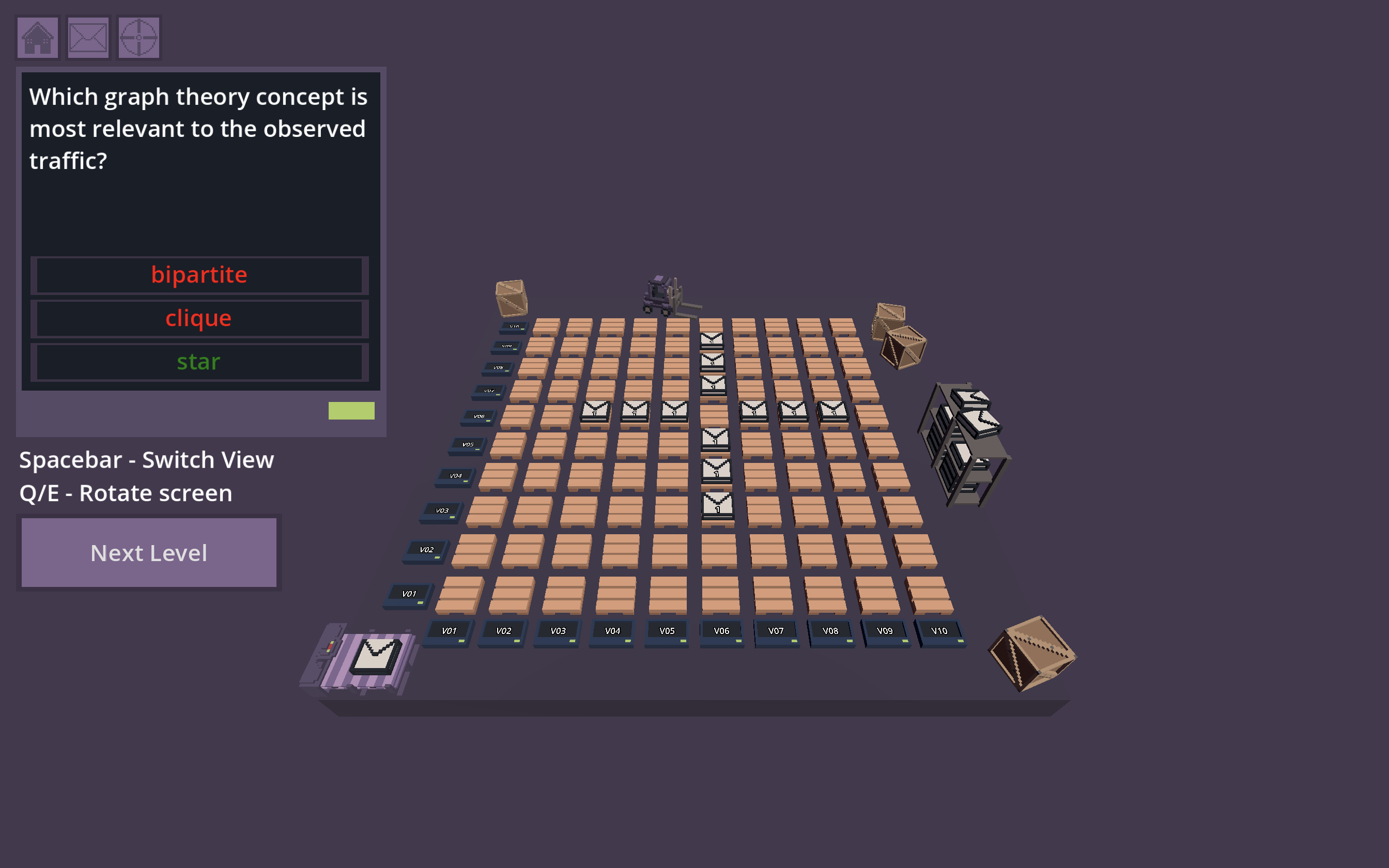}}
		\caption{Star.}
		\label{fig:gt-star}
	\end{subfigure}%
	~
	\begin{subfigure}{0.5\columnwidth}
		\center{\includegraphics[width=1.0\columnwidth,trim={0 5cm 10cm 0},clip]{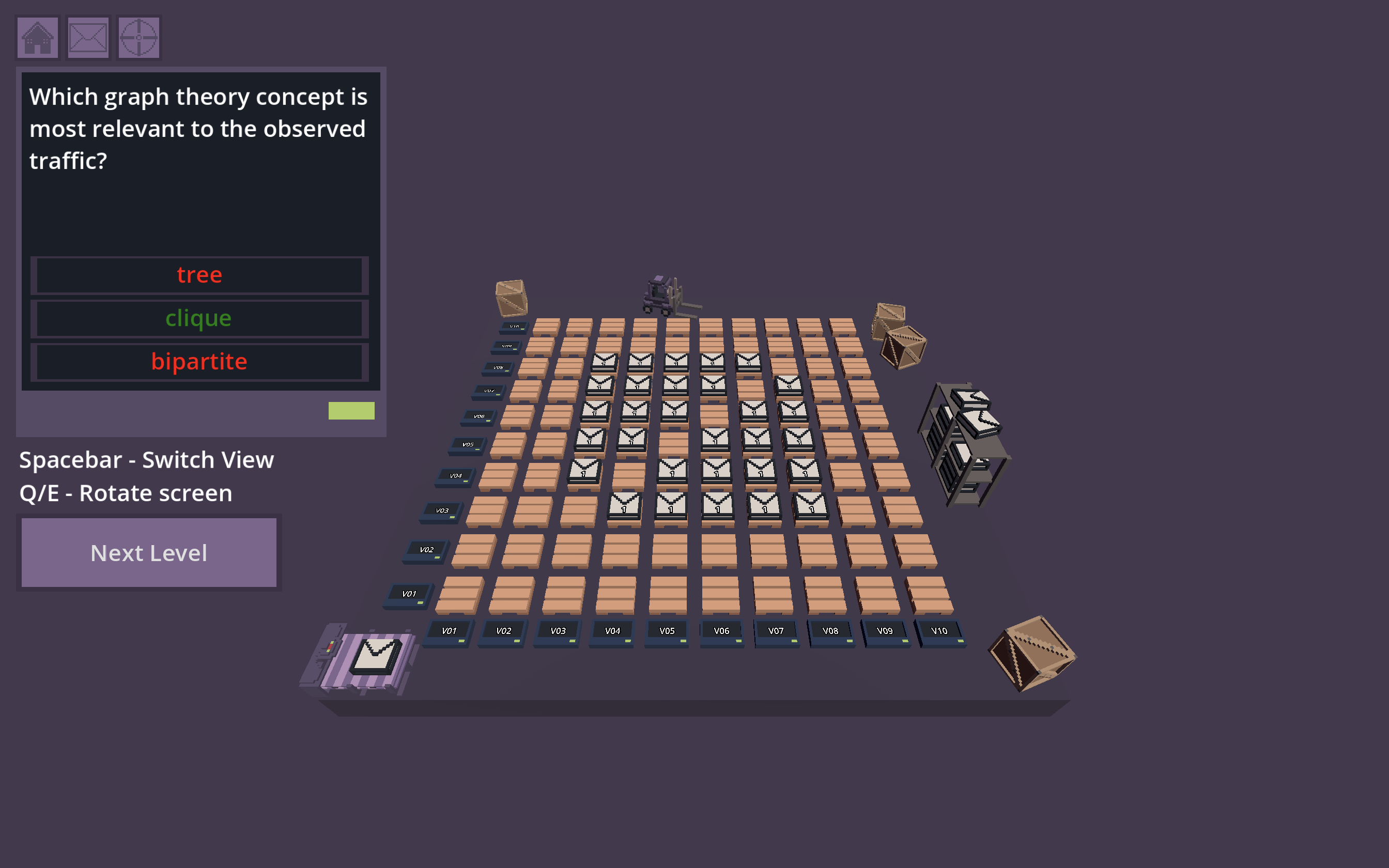}}
		\caption{Clique.}
		\label{fig:gt-clique}
	\end{subfigure}%
	\\
	\begin{subfigure}{0.5\columnwidth}
		\center{\includegraphics[width=1.0\columnwidth,trim={0 5cm 10cm 0},clip]{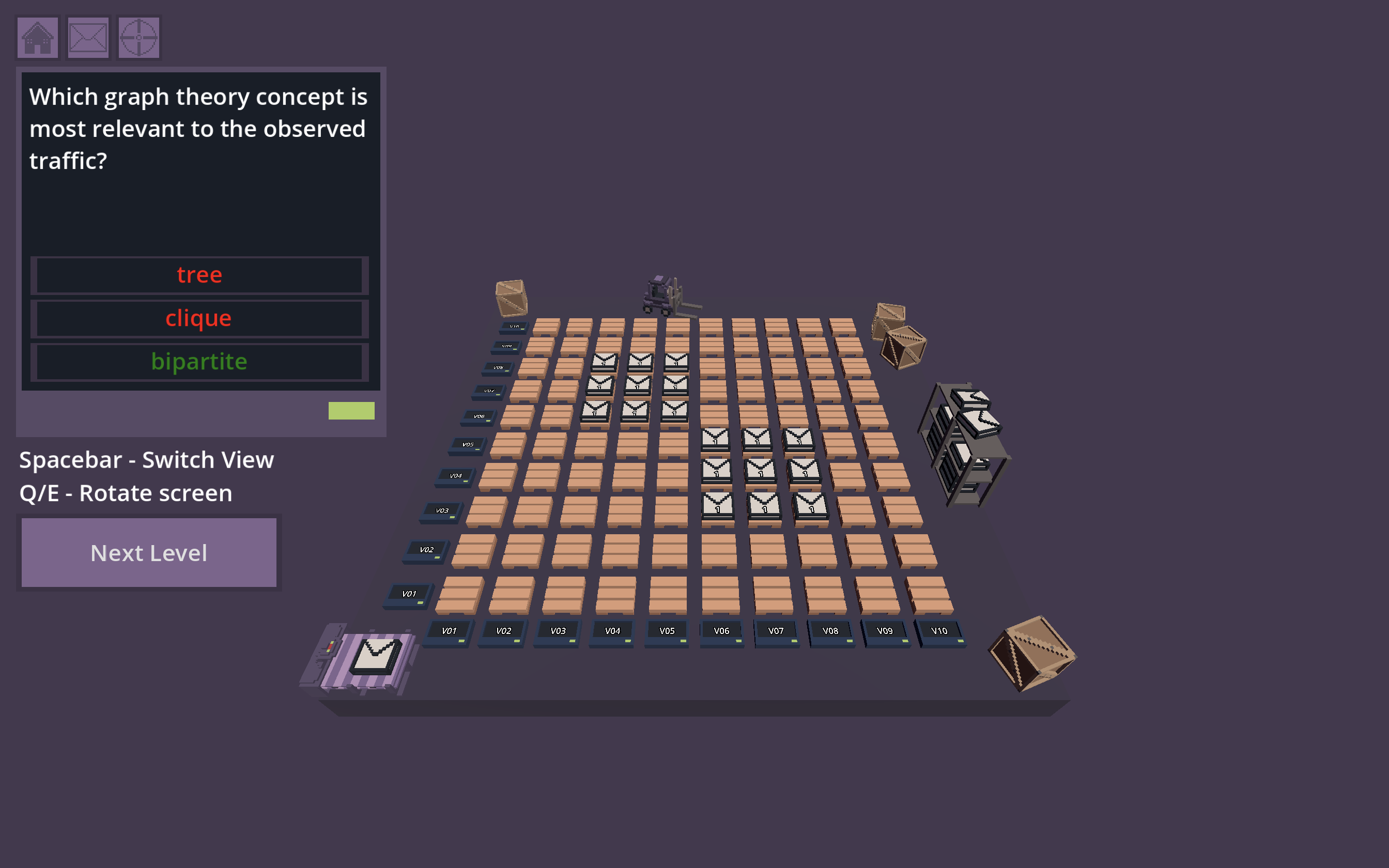}}
		\caption{Bipartite.}
		\label{fig:gt-bipartite}
	\end{subfigure}%
	~
	\begin{subfigure}{0.5\columnwidth}
		\center{\includegraphics[width=1.0\columnwidth,trim={0 5cm 10cm 0},clip]{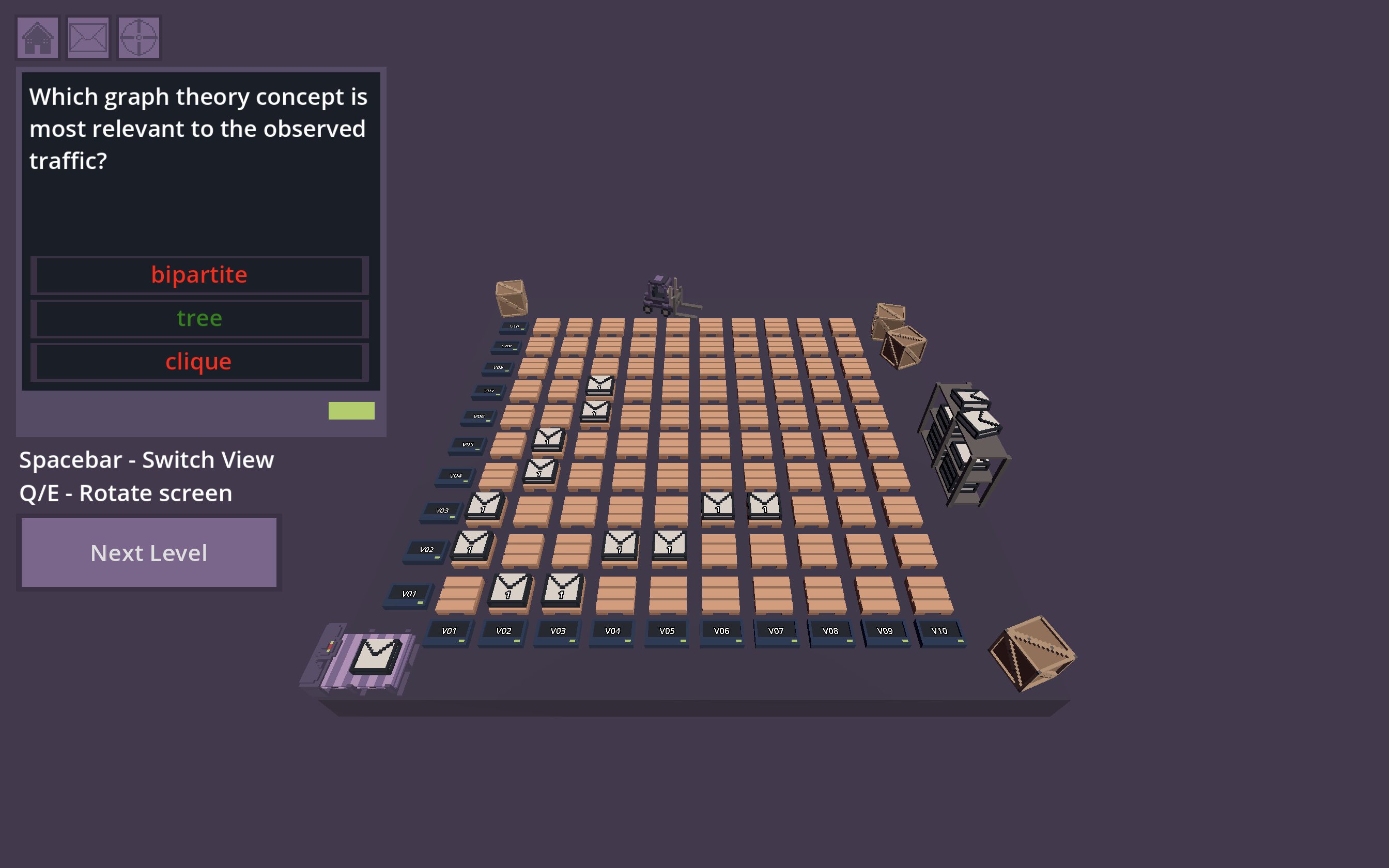}}
		\caption{Tree.}
		\label{fig:gt-tree}
	\end{subfigure}%
	\\
	\begin{subfigure}{0.5\columnwidth}
		\center{\includegraphics[width=1.0\columnwidth,trim={0 5cm 10cm 0},clip]{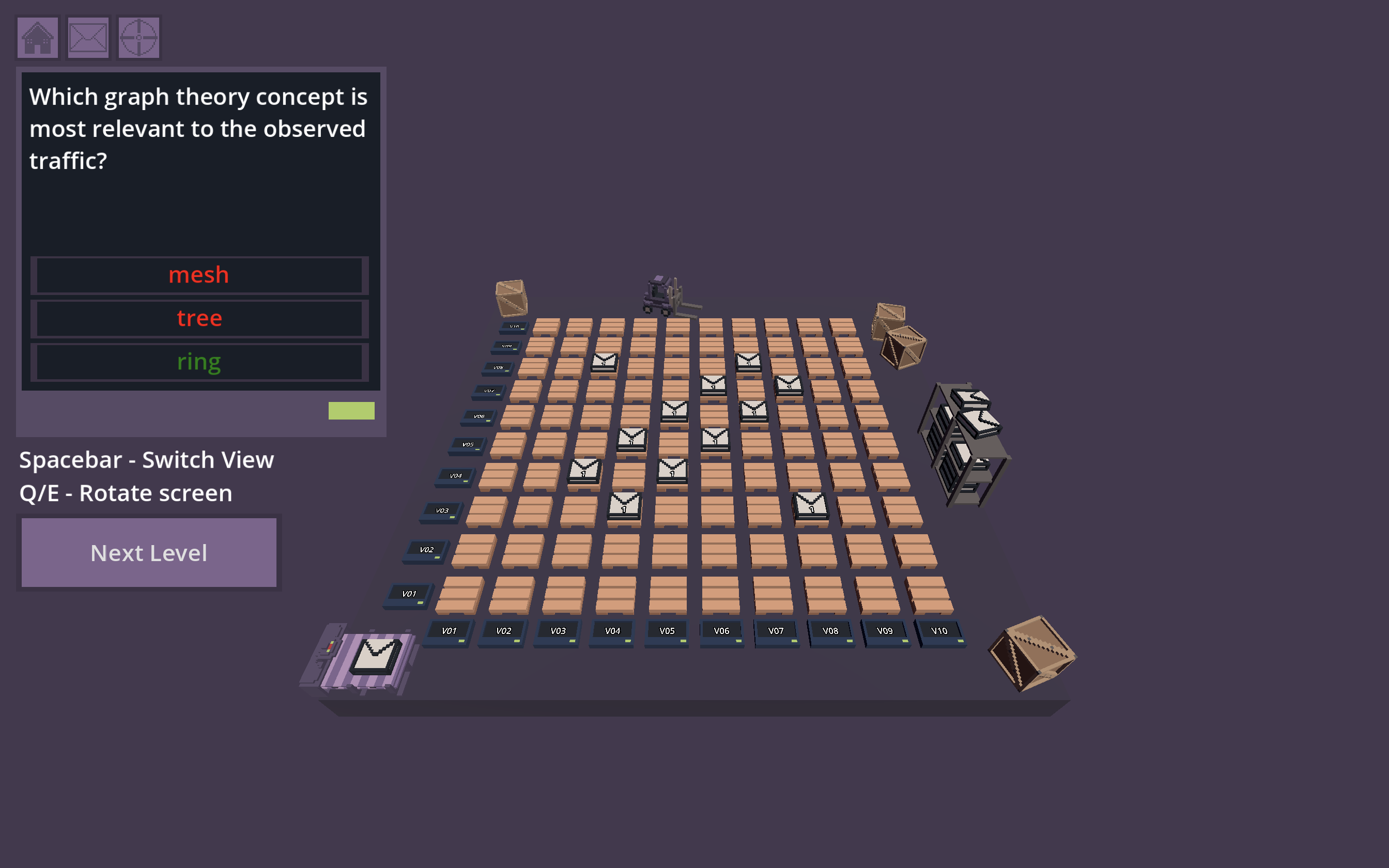}}
		\caption{Ring.}
		\label{fig:gt-ring}
	\end{subfigure}%
	~
	\begin{subfigure}{0.5\columnwidth}
		\center{\includegraphics[width=1.0\columnwidth,trim={0 5cm 10cm 0},clip]{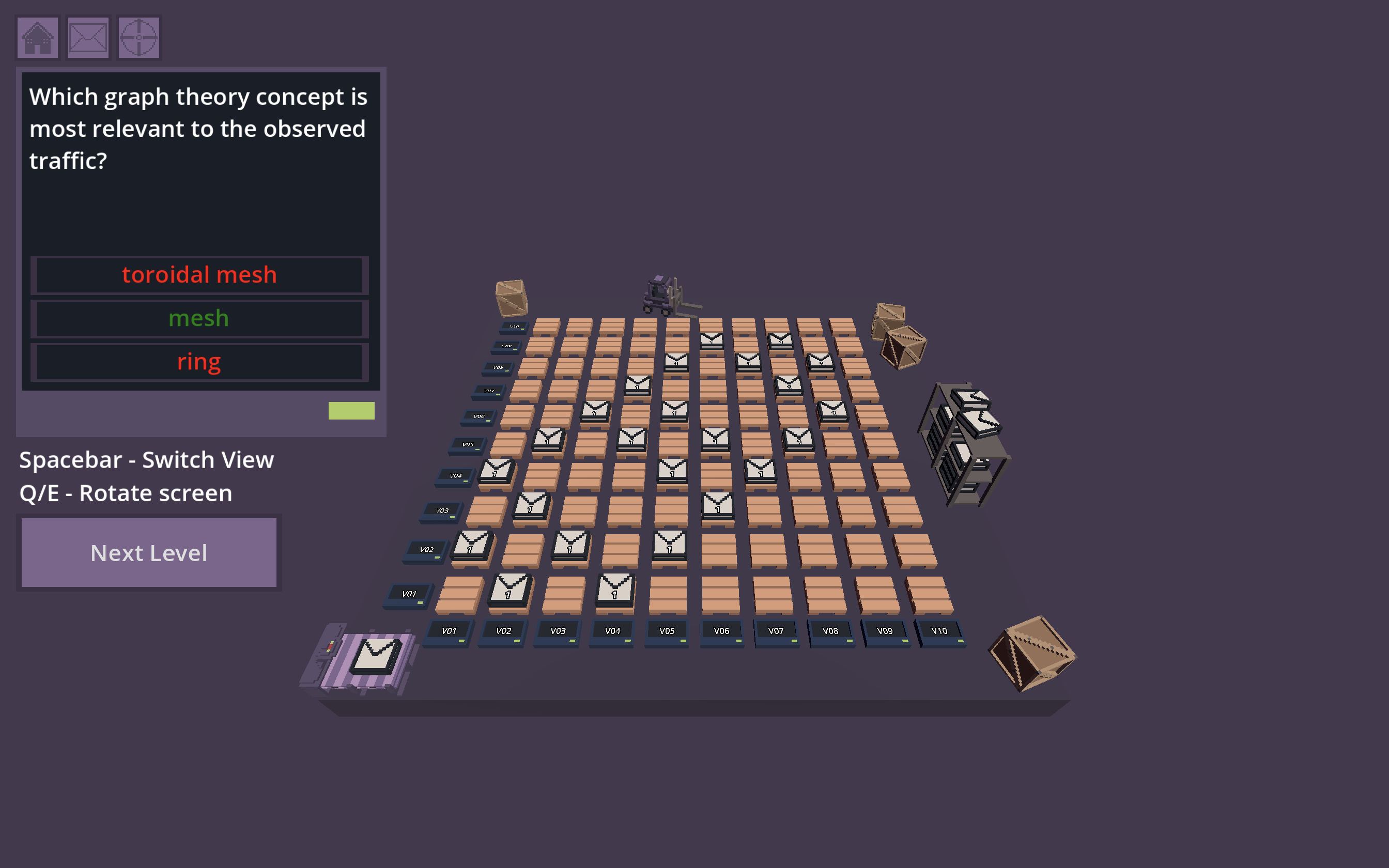}}
		\caption{Mesh.}
		\label{fig:gt-mesh}
	\end{subfigure}%
	\\
	\begin{subfigure}{0.5\columnwidth}
		\center{\includegraphics[width=1.0\columnwidth,trim={0 5cm 10cm 0},clip]{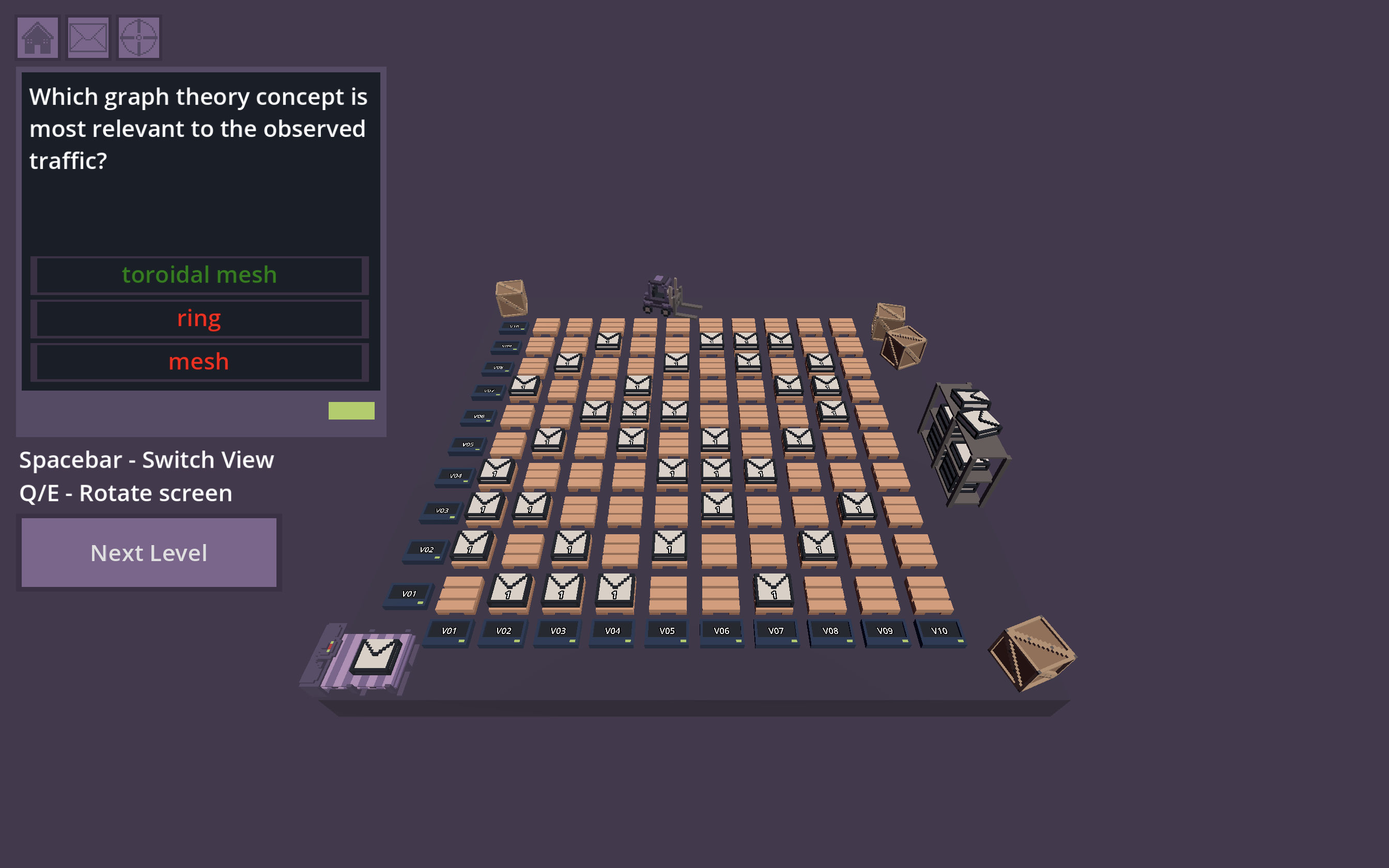}}
		\caption{Toroidal Mesh.}
		\label{fig:gt-toro-mesh}
	\end{subfigure}%
	~
	\begin{subfigure}{0.5\columnwidth}
		\center{\includegraphics[width=1.0\columnwidth,trim={0 5cm 10cm 0},clip]{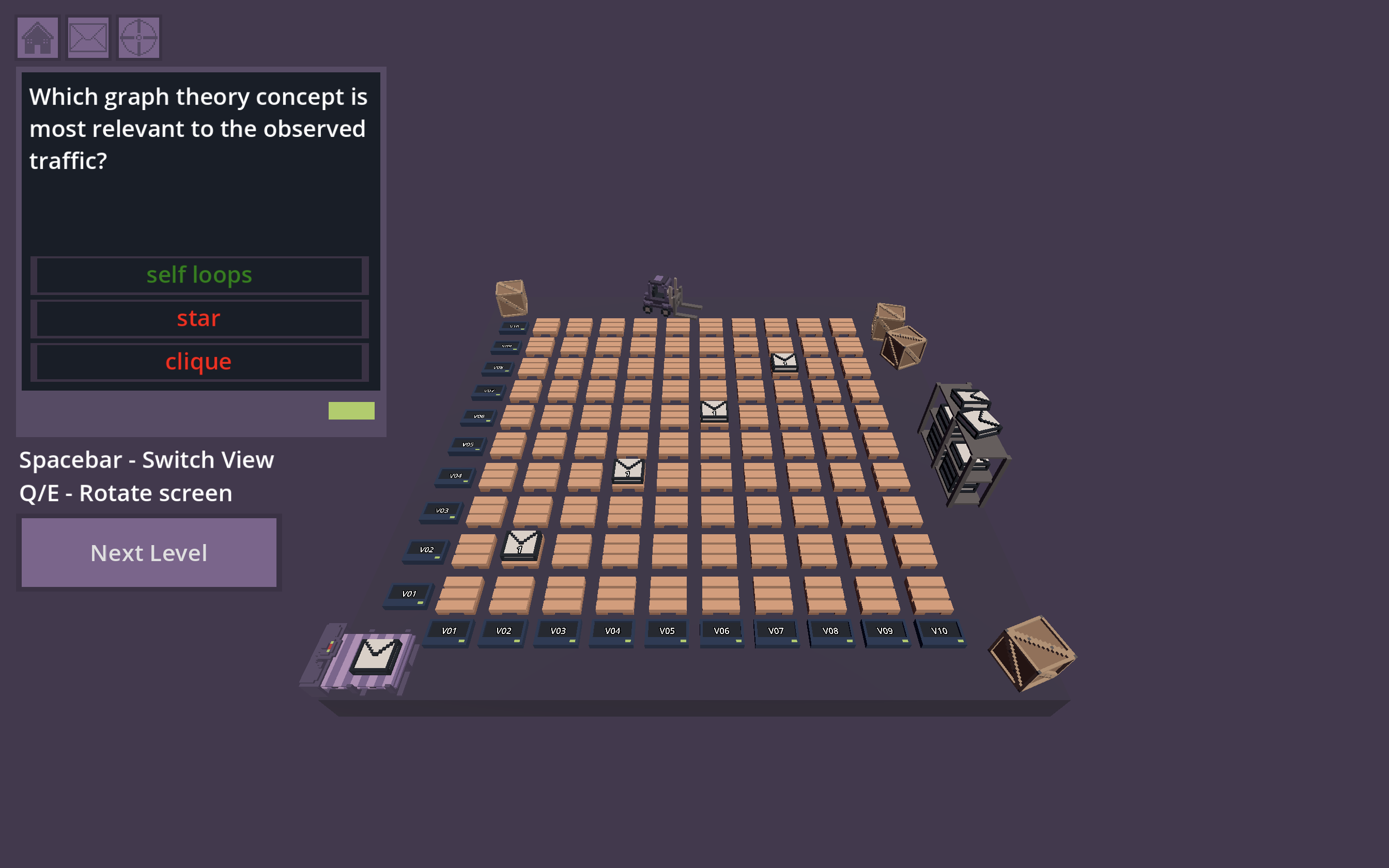}}
		\caption{Self Loop.}
		\label{fig:gt-loop}
	\end{subfigure}%
	\\
	\begin{subfigure}{0.5\columnwidth}
		\center{\includegraphics[width=1.0\columnwidth,trim={0 5cm 10cm 0},clip]{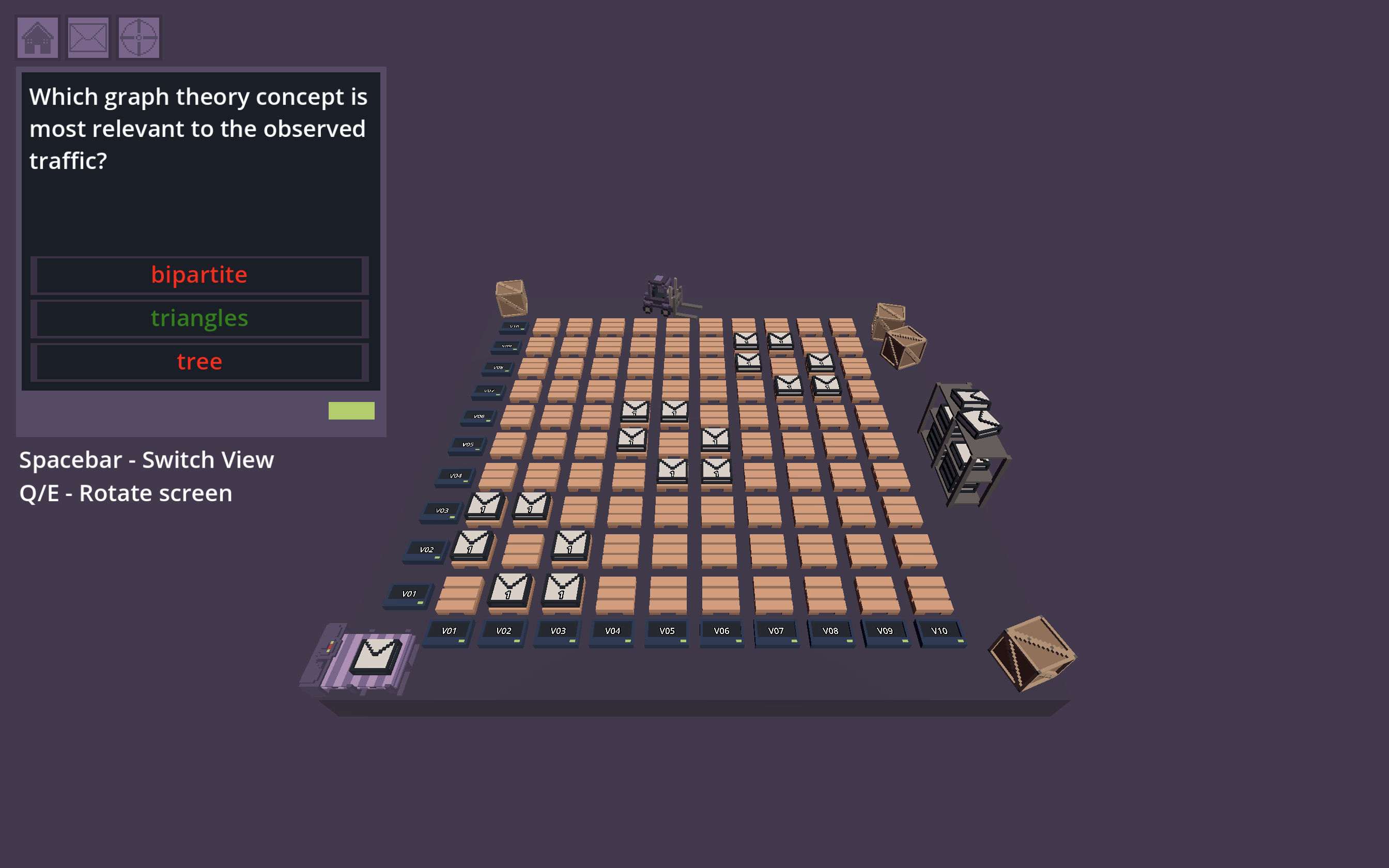}}
		\caption{Triangle.}
		\label{fig:gt-tri}
	\end{subfigure}
	\caption{{\bf Graph Theory.}  Traffic patterns relevant to various concepts in graph theory shown on a $10{\times}10$ traffic matrix.}
	\label{fig:graph-theory}
\end{figure}

The goal of \emph{Traffic Warehouse} is to allow the student to quickly engage with the learning modules in an intuitive manner.  When the student starts the game they are first shown a network traffic matrix in a top-down 2D view. This view is how they would generally see a matrix in a spreadsheet, a textbook, or a presentation, which allows the student to connect back to those learning modes and relate them to what they see on the screen. The student has the ability to go into a 3D mode by pressing the spacebar key.  The student can rotate the view using the Q and E keys. Switching between 2D and 3D and rotating the view allows the student to explore the network traffic matrix in new ways.

There is a single built-in module in \emph{Traffic Warehouse} and that is the training level (Figure~\ref{fig:matrix-training}). This module walks the player through what a traffic matrix is, how to read one, how it is of value to them, and how it will be represented in the game environment.  The training module also provides a space for the player to learn the controls of the game without needing to load in a learning module.  Because of the flexible nature of the JSON module file, the potential number of modules is very large.  Here a few will be described that illustrate the capabilities of  \emph{Traffic Warehouse}.   For all the modules, the question type is the same ``Which choice is the displayed traffic pattern most relevant to?''  Other questions types are possible.  This question type focuses on developing intuition in the student so they can connect the traffic patterns they are observing in a traffic matrix with common behaviors.  Often a hint is provided that directs the student to an external resource which may help them with the question.

The basic traffic topologies module presents traffic patterns shown for isolated links, single links, internal supernodes, and external supernodes with additional color coding to help provide context for these patterns (Figures~\ref{fig:tp-isolated-link} through \ref{fig:tp-ext-supe}, respectively).  These traffic topologies can help a student gain further understanding of the relationship between the source and destination pairs on the traffic matrix and what it actually means when there is a connection between the two.  After fully understanding the different topologies a student can move on to more advanced topics such as trying to analyze what is happening on a given matrix.

The notional attack module illustrates some of the common stages in a generic cyber attack.  Each individual stage of a cyber attack can easily be demonstrated on a traffic matrix. First is the planning stage, which is done in adversarial space (Figure~\ref{fig:na-plan}).  Second is staging, which takes place in greyspace (Figure~\ref{fig:na-stage}). Third is the infiltration stage, which happens at the border between grey and blue space (Figure~\ref{fig:na-infiltrate}).  The final stage is lateral movement, which happens inside blue space (Figure~\ref{fig:na-lat-move}).  The individual stages can be represented as they are (with only the traffic associated with each stage shown) in an effort to explain how the traffic pattern would look like in a real network.  After learning what the patterns look like individually, they could all be combined together or potentially mixed in with random background noise for a student to analyze and determine what is happening in the network.

A key concept in the protection of any domain is the distinction between (walls-in) security, (walls-out) defense, and deterrence \cite{kepner2021zero}.  The corresponding module illustrates these more abstract  concepts that are not associated with a specific attack, but places the broader idea of security, defense, and deterrence in a network traffic context.  For security, the student can see that the traffic is operating within their own blue space, communicating with their own systems and ensuring no adversarial activity (Figure~\ref{fig:sdd-sec}).  For defense, the student takes a step outside of their own network (Figure~\ref{fig:sdd-def}). This illustrates that  with community support from external sources, the student can identify threats to thier network before they have the chance to enter their own secured space.  Finally, deterrence is illustrated by showing that there is a credible activity in adversary space which arose as a response to unacceptable actions taken within the student's own network (Figure~\ref{fig:sdd-deter}).

Distributed denial-of-service (DDoS) attacks are amongst the most prevalent cyber attacks.  The DDoS module illustrates some of the more common components of a DDoS attack from a botnet.   Botnet command and control (C2) is shown by representing the communications in red space (Figure~\ref{fig:ddos-c2}). The communication from the C2 servers to the individual clients can be represented by identical communications between the C2 nodes and the botnet clients (Figure~\ref{fig:ddos-bot-clients}). The attack is then represented by communication from the botnet clients to the blue controlled servers (Figure~\ref{fig:ddos-attack}), followed by the backscatter when the servers reply back to the illegitimate traffic (Figure~\ref{fig:ddos-backscatter}).  Like other learning modules with multiple steps, after understanding these individual examples they could all be combined together or have background noise added to give a student even more of a challenge.

As a demonstration of the general flexibility of the \emph{Traffic Warehouse} JSON file, a module was created to illustrate a wide range of graph theory concepts. In graph theory graphs are mathematical structures used to model pairwise relations between objects.  Since the traffic matrix is simply a matrix filled with connections between two points it can represent different graphs in a way that a student may not have thought about which can help them further understand the concepts. This module demonstrates star, clique, bipartite, tree, ring, mesh, toroidal mesh, self loops, and triangle graphs (Figures~\ref{fig:gt-star} through \ref{fig:gt-tri}, respectively). The goal of providing these examples is to show that the information that can be displayed in \emph{Traffic Warehouse} is not limited just to network communication in the cybersecurity community.

\section{Conclusions and Future Work}

Network defense is critical in effectively maintaining an organization. Providing the best education available to our network security experts serves to increase their ability to keep our infrastructure and data safe. Using educational gaming environments can be a powerful way to take complicated or abstract concepts and teach them in a simple and intuitive way that most will be able to understand. In addition to educating practitioners, these educational gaming environments can be used as demonstrations in presentations to help non-technical participants get an understanding of what is being explained.  To fill this gap, the \emph{Traffic Warehouse} interactive game environment has been developed to teach the foundations of  traffic matrices to the computer networking community.  The game environment provides a convenient, broadly accessible, delivery mechanism that enables making material available rapidly to a wide audience.  The core architecture of the game is a facility to add new network traffic matrix training modules via an easily editable JSON file.  Using this facility an initial set of modules were rapidly created.

As with many educational or game development efforts, the potential for work and improvements is significant.  These range from lower level improvements (e.g., expanding the range of colors and materials for various objects, hierarchical learning modules, and obfuscating question answers in the module file) to broader work on any number of concepts in the cybersecurity domain such as firewall configuration, reverse engineering, access control, registry or filesystem structure, and more.  Furthermore, additional work should be done to find a rapid method of integrating educational games into already prepared course material and measuring the outcome and effect on the student.

\section*{Acknowledgments}

The authors wish to acknowledge the following individuals for their contributions and support:  Daniel Andersen, William Arcand, Sean Atkins, David Bestor, William Bergeron, Chris Birardi, Bob Bond, Koley Borchard, Stephen Buckley, Chansup Byun, K Claffy, Cary Conrad, Timothy Davis, Chris Demchak, Garry Floyd, Jeff Gottschalk, Daniel Grant, Dhruv Gupta, Chris Hill, Michael Houle, Matthew Hubbell, Charles Leiserson, Kirsten Malvey, Lauren Milechin, Sanjeev Mohindra, Guillermo Morales, Andrew Morris, Julie Mullen, Heidi Perry, Christian Prothmann, Andrew Prout, Steve Rejto, Albert Reuther, Josh Rountree, Antonio Rosa, Daniela Rus, Mark Sherman, Scott Weed, Charles Yee, Marc Zissman.

\bibliographystyle{ieeetr}
\bibliography{teaching-traffic-matrices}



\end{document}